\documentclass[fleqn,usenatbib]{mnras}

\usepackage{newtxtext,newtxmath}

\usepackage[T1]{fontenc}

\DeclareRobustCommand{\VAN}[3]{#2}
\let\VANthebibliography\thebibliography
\def\thebibliography{\DeclareRobustCommand{\VAN}[3]{##3}\VANthebibliography}

\usepackage{graphicx}	
\usepackage{amsmath}	
\usepackage{siunitx}

\newcommand\T{\rule{0pt}{2.6ex}}       
\newcommand\B{\rule[-1.2ex]{0pt}{0pt}} 
\newcommand{\minone}{$^{-1}$}
\newcommand{\tenpow}[1]{$10^{#1}$}

\newcommand{\hl}[1]{{\color{black}#1}}

\title[Snow lines stabilised in 2D using \textsc{cuDisc}]{\hl{CO} snow lines are stabilised by \hl{the vertical transport of volatiles}}

\author[A. Robinson et al.]{
Alfie Robinson,$^{1}$\thanks{E-mail: a.robinson21@imperial.ac.uk}
James E. Owen$^{1}$ and
Richard A. Booth$^{2}$
\\
$^{1}$Astrophysics Group, Imperial College London, Prince Consort Road, London SW7 2AZ, UK\\
$^{2}$School of Physics and Astronomy, University of Leeds, Leeds, LS2 9JT, UK
}

\date{Accepted XXX. Received YYY; in original form ZZZ}

\pubyear{2015}

\begin{document}
\label{firstpage}
\pagerange{\pageref{firstpage}--\pageref{lastpage}}
\maketitle

\begin{abstract}
Volatile evolution in protoplanetary discs determines the compositional evolution of forming planets. Below their sublimation temperatures, volatiles freeze out from the vapour phase onto dust grains in the disc and transition to being dynamically-coupled to the dust component as opposed to the gas. The boundary between the ice and vapour phases is referred to as the snow line, when thought of as the mid-plane radius at which the phase transition occurs, or the snow surface, when viewed as a 2D (radial and vertical) structure in the disc. We investigate whether the CO snow line (and therefore snow surface) is thermally unstable and therefore liable to changes in its location during disc evolution using the disc evolution code \textsc{cuDisc}, to which we have added an ice-vapour chemistry solver. We find that the instability does lead to there being two steady-state stable equilibrium solutions for the snow surface when including the vertical structure. However, in dynamically-evolving simulations, the disc does not enter a limit-cycle --- as seen in previous 1D models --- due to the \hl{shape of} the \hl{2D} snow surface and the vertical transport of volatiles. We therefore expect that dynamically evolution of snow lines due to instability is limited to transient, stochastic events rather than oscillatory behaviour with a regular period. However, we also expect the snow surface to evolve substantially during the disc lifetime solely due to changes in the thermal structure driven by evolution of the dust spatial structure and grain-size distribution --- this we will explore in future models.
\end{abstract}




\section{Introduction}

The initial composition of young planets is set by the solid and gaseous material that they accrete whilst forming in their host protoplanetary discs. In these discs, most of the important chemical elements are found in the form of simple molecules that are volatile at the temperatures and pressures found in such environments --- e.g. H$_2$O, CO$_2$, CO, N$_2$ and CH$_4$, amongst others \cite[see][for recent reviews of disc chemistry]{oberg2023,zhang2024}. These molecules can therefore either be in the vapour  phase or frozen-out onto the surfaces of dust grains at a certain location in the disc; the contour that marks the transition between the vapour and ice phase is referred to as the snow surface whilst the mid-plane radial location of this surface is called the snow line. The location of the snow line and the shape of the snow surface are set by the temperature structure of the disc, the binding energy of the volatile to the grain surfaces and the vapour pressure of the volatile. 

Understanding these features of discs is vital for explaining planet formation. As snow lines lie at different radii for each volatile, the compositions of the gas and dust components of the disc --- and therefore important quantities such as the C/O ratio --- are functions of disc location \citep{oberg2011}. Also, the dynamics of the ice and vapour phase volatiles are very different, as the ice-phase volatiles are coupled to the radially-drifting dust rather than the slowly-evolving gas. Such a difference means that composition is also a function of time \cite[see e.g.][]{stevenson1988,cuzzi2004,booth17,booth2019}. We also know that CO is depleted compared to the ISM abundance in many discs, primarily inferred from CO gas masses that suggest dust-to-gas ratios larger than those of the ISM \citep{ansdell2016,miotello2017}. Inclusion of dust growth and dynamics alongside chemistry when modelling the CO snow line \citep[see][]{kama2016,booth17,krijt2018,krijt2020,booth2019,powell2022,vanclepper2022} has indicated that these processes play a crucial role in CO depletion. More recently, JWST observations have suggested that low-mass stars ($<0.3~\mathrm{M}_\odot$) have high C/O ratios in the inner disc \citep{tabone2023,arabhavi2024,kanwar2024,long2025} in contrast to seemingly oxygen-rich higher mass stars \cite[see e.g.][]{romeromirza2024,temmink2024}. Insight into this differentiation requires comprehensive modelling of both chemistry and dust processes; using 1D models that combine volatile molecule chemistry with dust and gas transport, \cite{sellek2025a,sellek2025b} show that the C/O ratio in the inner disc regions is a time-dependent function of both chemical and dynamical timescales.

To understand the spatial structure, abundance and temporal evolution of molecules in discs, we therefore evidently need models of dust evolution that are coupled to the disc chemistry. CO in particular is an important molecule to study. It has a high interstellar abundance \citep{asplund2009}, both making it very important for outer disc chemistry and amenable to observation across multiple isotopologues (e.g. $^{12}$CO, $^{13}$CO, C$^{18}$O, C$^{17}$O; see \citealt{miotello2014,ansdell2016,bootha2019,law2023}). For typical discs around solar-mass stars, the mid-plane CO snow line lies in the outer disc, at radii of $>\sim20$~au \cite[see e.g.][]{oberg2011}. This means that it lies in a region of the disc where the temperature is governed by absorption and re-processing of stellar light by dust grains. In the aforementioned models, the disc thermal structure is not self-consistently coupled to the evolving dust and volatile distribution, and is usually held constant. As the CO snow line (and snow surface) is exponentially sensitive to the disc thermal structure, any evolution of the disc temperature due to changes in the dust population should have profound impacts on the molecular layout of the disc. As a case in point, \cite{owen2020} produced a model of CO evolution that included dust dynamics and coupled the disc thermal structure to the dust and CO ice distributions; their results suggested that the snow line can be unstable to thermal perturbations, leading to rapid radial movement of the mid-plane CO snow line. All of this points to the case that self-consistent modelling of the disc temperature structure alongside dust evolution and chemistry is vital for correctly modelling the evolution of molecules in discs.

In this paper, we study the CO snow line --- and snow surface --- in two dimensions (radial and vertical), coupling the disc thermodynamics to the time-dependent evolution of the dust and ice in the disc. To do this, we use the protoplanetary disc evolution code \textsc{cuDisc} \citep{robinson2024}, adding to the base model ice-vapour chemistry and ice opacities, all of which allows us to study the CO snow line in a thermodynamically-coupled, vertically-resolved disc. With this updated model, we build upon the work done studying the CO snow line instability in 1D by \cite{owen2020} by including the vertical dimension and a detailed model of the evolving grain-size distribution; in this paper, we discuss the effect that these inclusions have on the nature of the snow line instability. 

\section{Methods}

 We use \textsc{cuDisc} \citep{robinson2024}, a protoplanetary disc code that simulates dust evolution in the 2D $R-Z$ plane for axisymmetric discs and couples this evolving dust population to the thermal structure of the disc through solving the equations of radiative transfer using wavelength- and grain-size-dependent opacities. To study volatile evolution and study the movement of snow lines, we update \textsc{cuDisc} to include ice-vapour chemistry.

\subsection{Ice-vapour chemistry}\label{icevapourchemistry}

The ice-vapour chemistry model is based on those detailed by \cite{booth17,owen2020}, which in turn were based on \cite{hollenbach2009}. We consider adsorption and desorption of individual molecules from the surfaces of dust grains and construct rates for each of these processes. With this construction, we solve the following equations for the mass density of vapour and ice,

\begin{equation}
    \label{rhovap}
    \frac{\partial \rho_\text{v}}{\partial t} = \sum_k \left(\mathcal{R}_{\text{th},k} +\mathcal{R}_{\text{ph},k} - \mathcal{R}_{\text{ad},k}\right),
\end{equation}

\begin{equation}
    \label{rhoice}
    \frac{\partial \rho_{\text{i},k}}{\partial t} = \mathcal{R}_{\text{ad},k}- \mathcal{R}_{\text{th},k}-\mathcal{R}_{\text{ph},k},
\end{equation}
where $\rho_\text{v}$ is the mass density of the vapour, $\rho_{\text{i},k}$ is the mass density of  the ice condensed onto dust grains in size-bin $k$, and the rates are volatile adsorption, $\mathcal{R}_{\text{ad},k}$ and thermal desorption, $\mathcal{R}_{\text{th},k}$. The rate of volatile adsorption onto the grains is given by
\begin{equation}
    \label{Ra}
    \mathcal{R}_{\text{ad},k} = \pi a_k^2 v_\text{th} n_k \rho_\text{v},
\end{equation}
where $a_k$ and $n_k$ are the radius and number density of grain size $k$, and $v_\text{th}$ is the thermal velocity, 

\begin{equation}
    \label{vtherm}
    v_\text{th} = \sqrt{\frac{8 k_B T}{\pi m_\text{mol}}},
\end{equation}
$m_\text{mol}$ being the mass of the volatile molecule. The rate of thermal desorption is given by the Polanyi-Wigner equation \citep{bergin2007},

\begin{equation}
    \label{numlayers}
    \mathcal{R}_{\text{th},k} = 4\pi a_k^2 N_s n_k m_\text{mol} p_{k} \nu_i \exp\left(-\frac{T_\text{bind}}{T}\right),
\end{equation}
where $N_s$ is the number of adsorption sites per unit area, set by default to \num{1.5e15}~\unit{\per\cm\squared} \citep{aikawa1996}, $T_\text{bind}$ is the temperature of the binding energy of the molecule (i.e. $E_\text{bind}=k_BT_\text{bind}$ --- such binding energies are derived from desorption rates measured in Temperature Programmed Desorption (TPD) experiments; see \citealt{Cuppen2017}) and $\nu_i$ its vibrational frequency, estimated by

\begin{equation}
    \label{nu_i}
    \nu_i = \sqrt{\frac{2 N_s k_BT_\text{bind}}{\pi^2 \mu_\text{mol}m_H}},
\end{equation}
where $\mu_\text{mol}$ is the molecular weight of the molecule. $p_{k}$ is the probability of an adsorption site having a molecule bonded to it. Here, we assume that adsorbing molecules distribute randomly across the grain surface, such that
\begin{equation}
    \label{p_k}
    p_{k} = 1-\exp\left( -\dfrac{\rho_{\text{i},k}}{4\pi a_k^2N_sn_km_{\text{mol}}} \right).
\end{equation}
This smoothly interpolates between the zeroth-order rate, when the grains are completely coated in volatiles and ${p_{k}=1}$, and the first-order rate, when the grains are only partially coated. A caveat of this method is that it ignores potential distributions in the molecule-to-grain binding energies, which have been observed in laboratory experiments and predicted by quantum chemistry calculations \cite[see][]{bovolenta2022,furuya2024}. We leave this as a topic for future models to explore. Some previous work has also shown that nucleation of water ice on bare silicate grains requires larger vapour pressures than that required for ice-coated grains \citep{ros2019}, however since these effects have not been studied for CO ice, we do not take them into account for the work presented in this paper. Likewise, we do not currently include the effect of the release of latent heat during sublimation, as for CO it has a negligible impact on heating compared to changes in opacity \cite[see][]{owen2020}\hl{; however, studies have shown it to be important when studying the water snow line \citep{lecar2006,wang2025}}.

Photodesorption of adsorbed volatile molecules is triggered by UV photons from both the star-disc system and FUV photons generated by cosmic rays. For these two processes, we follow \cite{Cuppen2017} and \cite{Sternberg1987,Hartquist1990} respectively, and arrive at the following photodesorption rate,

\begin{equation}
    \label{R_ph}
    \mathcal{R}_{\text{ph},k} = \pi a_k^2 n_k m_\text{mol} p_{k} \left(\gamma_\text{UV}+\gamma_\text{CR}\right) Y
\end{equation}
where $\gamma_\text{UV}$ and $\gamma_\text{CR}$ and the UV photon fluxes (photons~\unit{\per\cm\squared \per\second}) generated by the star-disc system and cosmic rays respectively and $Y$ is the photodesorption yield, taken to be \num{2.7e-3}~molecules~photon\minone \citep{oberg2009}. The photon flux from the star-disc system, $\gamma_\text{UV}$, is calculated by converting both the direct stellar radiation field and the radiative energy in each of the FLD wavelength-bands to a photon flux, taking energy only from bands with wavelengths less than 200~nm. The FUV flux stimulated by cosmic rays, $\gamma_\text{CR}$, is found by balancing the rate at which UV photons are produced by cosmic rays \citep{Sternberg1987} with the rate at which they are adsorbed by dust grains and is given by

\begin{equation}
    \label{gamma_CR}
    \gamma_\text{CR} = 0.15 \eta_\text{CR} \frac{n_{\text{H}_\text{nuc}}}{\sum_k \pi a_k^2 n_k},
\end{equation}
where $\eta_\text{CR}$ is the rate of cosmic ionisations per hydrogen nucleus, taken as \tenpow{-17}~\unit{\per\second}, and $n_{\text{H}_\text{nuc}}$ is the number density of hydrogen nuclei, calculated from the gas density assuming that it is solely composed of hydrogen and helium at ISM abundances. \hl{For the work presented here looking at the CO snow line, photodesorption has a negligible effect because thermal desorption constrains the snow surface to altitudes below those at which photodesorption becomes important.}
\begin{figure*}
    \centering
    \includegraphics[width=.99\textwidth,trim={0.5cm 0.4cm 0.4cm 0.cm},clip]{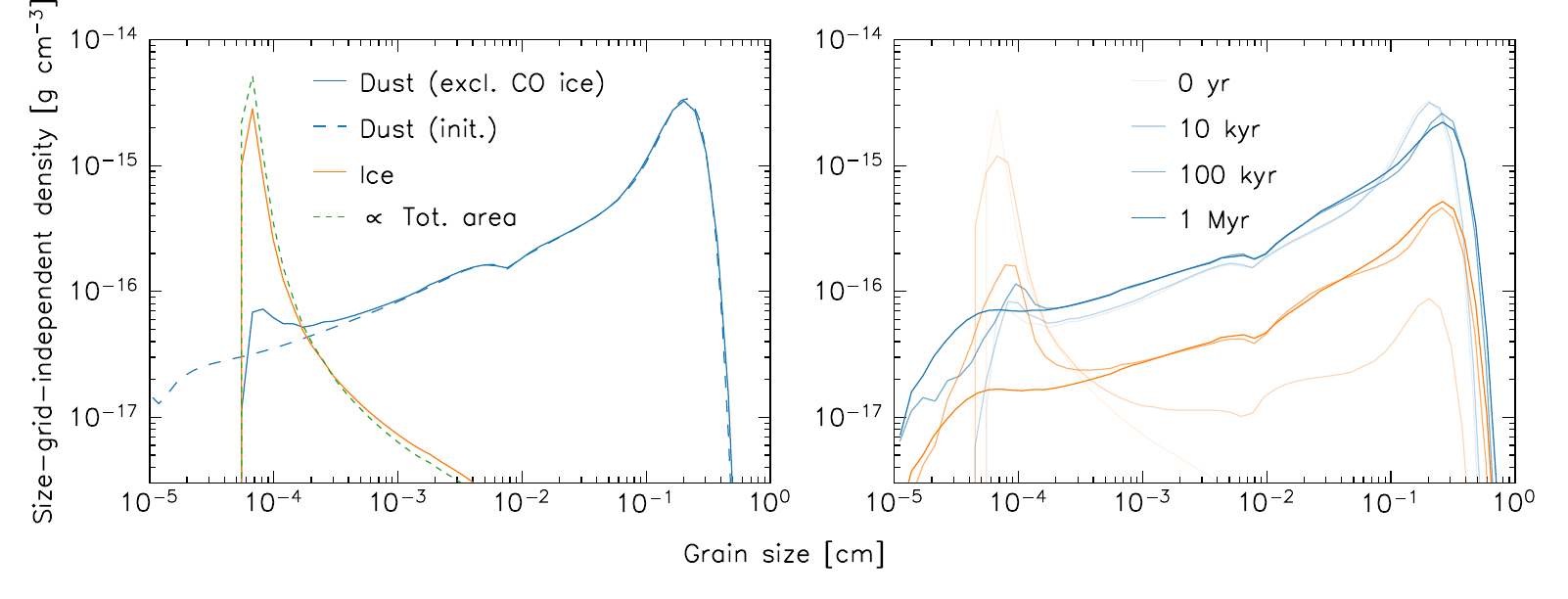}
    \caption[Ice distribution]{The left-hand panel shows grain-size-independent mass densities of ice and dust for vapour condensing onto a static distribution of dust grains. The dashed blue line shows the initial non-icy dust distribution, whilst the dashed green line is proportional to the total grain surface in a bin, i.e. $\propto n_ka_k^2$. The right-hand panel shows grain-size-independent mass densities of ice and dust evolving through collisions from the initial state shown in the left-hand panel until a steady-state is reached.}
    \label{icedistinit}
\end{figure*}

Eqns. \ref{rhovap}~\&~\ref{rhoice} are solved using a first-order iterative implicit method. First, we rewrite the equations in a form where the rates are linearly dependent on the vapour and ice densities, then discretise them implicitly:

\begin{equation}
    \label{rhovapdisc}
    \frac{\rho_\text{v}^{n+1}-\rho_\text{v}^n}{\Delta t} = \sum_k \left(D_{k} \rho_{\text{i},k}^{n+1} - A_{k}\rho_\text{v}^{n+1}\right),
\end{equation}

\begin{equation}
    \label{rhoicedisc}
    \frac{\rho_{\text{i},k}^{n+1}-\rho_{\text{i},k}^{n}}{\Delta t} = A_{k} \rho_\text{v}^{n+1} - D_{k}\rho_{\text{i},k}^{n+1},
\end{equation}
where $A_{k}$ and $D_{k}$ are the adsorption and desorption rates in units of molecules~\unit{\per\second}. This system can be solved analytically, and updates to each of the vapour and ice species are calculated through the following steps:

\begin{enumerate}
    \label{rhovapdisc}
    \item $\tilde{\rho}_\text{i} \equiv \displaystyle\sum\limits_k \dfrac{D_k\Delta t \rho_{\text{i},k}^n}{1+D_k\Delta t}$
    \item $\tilde{A} \equiv \displaystyle\sum\limits_k \dfrac{A_k\Delta t}{1+D_k\Delta t}$
    \item $\rho_\text{v}^{n+1} = \dfrac{\rho_\text{v}^n + \tilde{\rho}_\text{i}}{1+\tilde{A}}$
    \item $\rho_{\text{i},k}^{n+1} = \dfrac{\rho_{\text{i},k}^n + A_k\Delta t \rho_\text{v}^{n+1}}{1+D_k\Delta t}$
\end{enumerate}
where $\tilde{\rho}_\text{i}$ and $\tilde{A}$ are intermediate sums. 

Since both $A_k$ and $D_k$ depend on the ice mass density implicitly through the grain radius, $a_k$, and the covering fraction, $f_{C,k}$, we iterate the above steps, recalculating the rates at the start of each iteration. Before commencing the next iteration, grain sizes and internal matter densities are recalculated to incorporate the new ice mass-fractions. Iterations are continued until convergence to a tolerance is achieved in the final vapour and ice mass densities. The tolerance is set through calculating the mean fractional error across all ice species and vapour over the entire spatial grid between subsequent iterations. The default tolerance is a mean fractional error of 10$^{-5}$. 

\begin{table}
    \centering
    \begin{tabular}{ccc}
        \hline Parameter & Symbol &  Value \T \B \\ \hline
        \T Viscosity & $\alpha$ & \tenpow{-4} \\ 
        Dust-to-gas ratio & - &  \tenpow{-2}\\
        CO ice internal density & $\rho_{m,\text{CO}}$ & 0.89 \unit{\gram\per\cm\cubed}\\
        Base dust internal density & $\rho_{m,\text{dust}}$ & 1.67 \unit{\gram\per\cm\cubed}\\
        Cylindrical radius & $R$ & 40 au \\
        Gas surface density & $\Sigma_g$ & 600$(R/\text{au})^{-1}$ \unit{\gram\per\cm\squared} \\
        Temperature & $T$ & 130$(R/\text{au})^{-\frac{1}{2}}$ K \\ 
        Fragmentation velocity & $v_\text{frag}$ & 100 \unit{\cm\per\second}\\
        CO binding energy temperature & $T_\text{bind,CO}$ & 850 K \\ \hline
    \end{tabular}
    \caption[Parameters used for the 0D and 1D simulations of ice-vapour chemistry and dynamics.]{Parameters used for the 0D and 1D simulations. The CO binding energy temperature is assumed to be that of pure CO ice, and is taken from \cite{bisschop2006}.}
    \label{tab:simprunparams}
\end{table}

The left-hand panel of Fig.~\ref{icedistinit} shows the equilibrium distribution of icy grains for CO vapour condensing onto a static distribution of dust grains, i.e., the equilibrium distribution when coagulation is neglected. The parameters for the model are shown in Table~\ref{tab:simprunparams}. When plotting the distribution, the dust and ice densities are re-binned onto a uniformly-spaced logarithmic mass-grid, and the grain size-grid is recalculated using the binned densities because several of the smallest mass bins end up with very similarly-sized grains. The binning is carried out in the same mass-conserving method used to distribute coagulation products into the correct bins --- see Eqn.~65 in \cite{robinson2024}. To avoid spurious noise in the binned distribution that arises due to periodicities between the two mass-grids, the re-binned grid uses just over half the number of simulation mass bins. \hl{As the internal density in each mass-bin can vary, the size-grid becomes non-uniform. To remove the effect this has on the plotted grain-size distribution, we plot the quantity $\frac{\rho_k}{\text{d}\ln a}$, i.e. the grain-size-grid-independent mass density, as done in works such as \cite{birnstiel2010}.}

The ice distribution is almost proportional to the total surface area of each grain size bin, $n_ka_k^2$, shown by the dashed green line. This occurs because the adsorption rate is proportional to this area, so in the initial stages of ice adsorption, when desorption rates are negligible in comparison to adsorption rates, the ice forms a distribution that is proportional to area. The distribution is not exactly proportional however, as the smallest grains have grown considerably compared to their initial size, so the initial surface area available for adsorption was smaller than the final area shown by the dashed green line.

\subsection{Ice growth}

In our growth model, the ice component of each grain is treated as a tracer of the non-ice base component. As grains collide and mass is transferred between mass bins, the ice is assumed to also be transferred proportionally according to the ice mass-fraction in each bin. The rate of ice density gain due to collisions, $\dot{\rho}_{\text{i,gain}}$, is found by multiplying the rate for the
associated non-ice component by the ice to non-ice density ratio,

\begin{equation}
    \label{coagloss_ice}
    \dot{\rho}_{\text{i,loss}, ij} = \left(\frac{\rho_{\text{i},j}}{\rho_{\text{d},j}}\right) m_j n_i n_j K_{ij} (p_{{\rm coag},ij} + p_{{\rm frag},ij}),
\end{equation}
where $\rho_{\text{d},j}$ is the density of the non-ice dust component, $m_i$ and $m_j$ are the masses of the non-ice components, and $K$, $p_{{\rm coag}}$, and $p_{{\rm frag}}$ are the coagulation kernel and the coagulation and fragmentation probabilities as given in \cite{robinson2024}. The rate of loss of ice density due to collisions, $\dot{\rho}_{\text{i,loss}}$, is then given by

\begin{multline}
    \dot{\rho}_{\text{i,gain}, ijk} = n_i n_j K_{ij} \left[p_{\text{coag},ij} M_{\text{i},ij} C_{ijk}\right.\\
    \left. + p_{\text{frag},ij} \bar{X}_{ij} \left(m_{\text{rem},ij}R_{ijk} + m_{\text{frag},ij}F_{ijk}\right)\right],
    \label{ngain_ice}
\end{multline}
where $C$, $R$, and $F$ are the collision product coefficients, and $m_{\text{rem}}$ and $m_{\text{frag}}$ are the remnant and fragment masses, again all defined in \cite{robinson2024}. $M_{\text{i}}$ is the total colliding ice mass,
\begin{equation}
    \label{m_ice_tot}
    M_{\text{i},ij} = \left(\frac{\rho_{\text{i},i}}{\rho_{\text{d},i}}\right)m_i + \left(\frac{\rho_{\text{i},j}}{\rho_{\text{d},j}}\right)m_j,
\end{equation}
and $\bar{X}$ is the total ice-to-dust ratio,
\begin{equation}
    \label{frag_frac_tot}
    \bar{X}_{ij} = \frac{\frac{\rho_{\text{i},i}}{n_i}+\frac{\rho_{\text{i},j}}{n_j}}{m_i+m_j}.
\end{equation}

This method makes the assumption that the ice and non-ice components are equally mixed throughout the grain; in reality, the ice predominantly forms a mantle around a non-ice core, however incorporating this physics currently goes beyond the scope of this work. Also, the inclusion of this piece of physics would only significantly impact fragmentation or abrasion, which are less important processes at the CO snow line due to radial drift. 

The Stokes numbers used to determine the grain relative velocities also depend on the ice mass-fraction through the grain size and internal matter density, which vary with the ice fraction of the grains. We keep the mass-grid used in the growth calculations fixed to the masses of the base dust grains.

The right-hand panel of Fig.~\ref{icedistinit} shows 0D growth evolution at the disc mid-plane with the same setup as for the pure adsorption-desorption calculation shown in the left-hand panel. As there is no dynamics, the dust distribution is limited by fragmentation rather than radial drift. We see that the ice becomes equally-distributed amongst the mass bins when in coagulation-fragmentation equilibrium. This is because, at equilibrium, the net condensation and evaporation rates are the same for all grains and do not depend on the ice mass-fraction. Hence, growth with fragmentation ultimately redistributes the ice across all grain sizes.

\subsection{Ice-vapour transport}\label{icevapdyn}

Ice that condenses onto grains in a certain dust mass-bin travels at the velocity of the grains in that bin. The drag forces that govern these velocities are calculated using the radius and density of the total grain (ice and dust). Using these velocities and the ice density in each bin, we calculate the total ice density-flux due to advection and diffusion using the same method described in \cite{robinson2024}. To conserve momentum, we update the total momentum of the dust and ice by summing the fluxes. The momentum of each component can then be updated using the appropriate mass-fraction. As the size and internal matter density of the grains in each cell are updated based on the local ice mass-fraction, the Stokes numbers are also correctly calculated using these values during the source step of the dynamics calculation. Advective and diffusive fluxes are also calculated for the vapour, assuming it travels at the gas velocity. 
\begin{figure*}
    \centering
    \includegraphics[width=.9\textwidth]{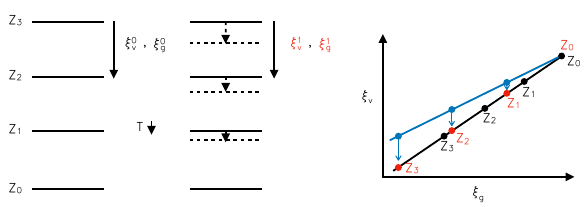}
    \caption[Cartoon demonstrating method for updating vapour column during hydrostatic equilibrium calculations]{Schematic showing the method used for updating the vapour column during hydrostatic equilibrium calculations. The left two cartoons show how the column is affected by a temperature decrease that causes the gas density to increase at lower altitudes. The dashed lines show the new altitudes of the column values that were initially at the heights $Z_0$ up to $Z_3$. The black line on the plot on the right shows the initial vapour column as a function of gas column, whilst the blue line shows the initial vapour column now as a function of the new gas column. At each $Z$-value, the gas column has decreased, and the vapour must follow suit. We then interpolate along the original vapour-gas distribution (black line) to the find the updated vapour column values that lie at the new gas column values, indicated by the red points.}
    \label{columncal}
\end{figure*}

\hl{We also adjust the vapour density during hydrostatic equilibrium calculations using a Lagrangian remapping approach: as the gas mass is redistributed vertically, we likewise redistribute the vapour mass to give the same vapour distribution as a function of vertical gas column density. In summary, the steps of the calculation are: 

\begin{enumerate}
    \item Perform the standard hydrostatic equilibrium update for the gas.
    \item Calculate the gas column as a function of height before and after the update, and calculate the initial vapour column.
    \item Interpolate to find the updated vapour column that reproduces the initial vapour column as a function of gas column, given the updated gas column.
\end{enumerate}}

The column of either vapour or gas is calculated through

\begin{equation}
    \xi(Z) = \ln\int_\infty^Z\rho \, \text{d}Z',
\end{equation}
where the integration is performed from the surface to the mid-plane to avoid losing accuracy near the surface due to the limits on floating-point precision. The gas column is calculated before and after the hydrostatic equilibrium update, and the vapour column is calculated similarly before the update. We now want to calculate new vapour column values so that the vapour column as a function of the gas column remains constant, i.e., the vapour has moved the same distance vertically as the gas. 

This is achieved by interpolating on the grid of original vapour and gas column values to calculate the new vapour column values that lie at the new gas column values, as demonstrated in Fig.~\ref{columncal}. This interpolation is given by

\begin{equation}
    \xi_\text{v}^{1}(Z_j) = \xi_\text{v}^0(Z_l)+\left[\xi_\text{v}^0(Z_{l-1})- \xi_\text{v}^0(Z_{l}) \right]\frac{\xi_\text{g}^1(Z_j)-\xi_\text{g}^0(Z_l)}{\xi_\text{g}^0(Z_{l-1})- \xi_\text{g}^0(Z_{l})},
\end{equation}
where the superscripts $0$ and $1$ indicate the original and new column values respectively, $Z_l$ is the altitude that satisfies $\xi_\text{g}^0(Z_l)<\xi_\text{g}^1(Z_j)<\xi_\text{g}^0(Z_{l-1})$. Note that $\xi$ increases with decreasing $Z$ since we are integrating from the surface down. The interpolation is performed in log-log space to increase accuracy given the quasi-exponential forms of the column densities. The final vapour densities are calculated via

\begin{equation}
    \rho^1_\text{v} = \frac{\text{d}\exp{(\xi^1_\text{v}})}{\text{d}Z}.
\end{equation}

\subsection{Thermodynamic coupling: CO ice opacities}

To couple the ice component of each grain to the thermodynamics of the system, we implement the opacity model detailed by \cite{cuzzi2014}. This model allows the calculation of absorption and scattering opacities on-the-fly, as opposed to the use of interpolated opacity tables. It is an approximation to full Mie theory \citep{mie1908}. Each grain is taken to be composed of a number of different species, each species $k$ with tabulated optical constants -- either a complex dielectric function, $\epsilon_k$, as a function of wavelength or complex refractive index, $N_k = \sqrt{\epsilon_k}$ -- and a mass-fraction of the total grain mass, $f_{m,k}$. By default, each grain has the same base composition as that used by the DSHARP model (\citealt{DSHARP2018}, see Table~\ref{tab:dsharpcomp})\footnote{Given that this paper focuses on the CO snow line, which resides in the outer disc at radii $>10$~au, we do not update the composition to remove the water ice component for dust interior to the location of the water snow line at $\sim1-5$~au; this is an addition that we would be interested in studying when investigating the inner disc in the future.}. In addition to this base composition, CO contributes to the overall grain composition with a mass-fraction given by $f_{m,\text{CO}} = \rho_\text{i}/(\rho_\text{d}+\rho_\text{i})$, where $\rho_\text{i}$ and $\rho_\text{d}$ are the CO ice density and dust density in a given spatial cell and base dust-grain-mass bin. 
\begin{table}
    \centering
    \begin{tabular}{ccc}
        \hline Species &  mass-fraction & Reference \T \B \\ \hline
        \T Organics & 0.397 & \cite{henning1996} \\ 
        Silicates &  0.329 & \cite{draine2003} \\
        Water ice & 0.2 & \cite{warren2008} \\
        FeS & 0.074 & \cite{henning1996} \\ \hline
    \end{tabular}
    \caption[The base composition used for dust grains in the mixed-composition ice and dust opacity model.]{The base composition used for dust grains, taken from the DSHARP model \citep{DSHARP2018}. The references from which the optical constants have been taken are given for each species.}
    \label{tab:dsharpcomp}
\end{table}

\begin{figure*}
    \centering
    \includegraphics[width=.98\textwidth, trim={0.5cm 0.45cm .5cm 0.45cm}, clip]{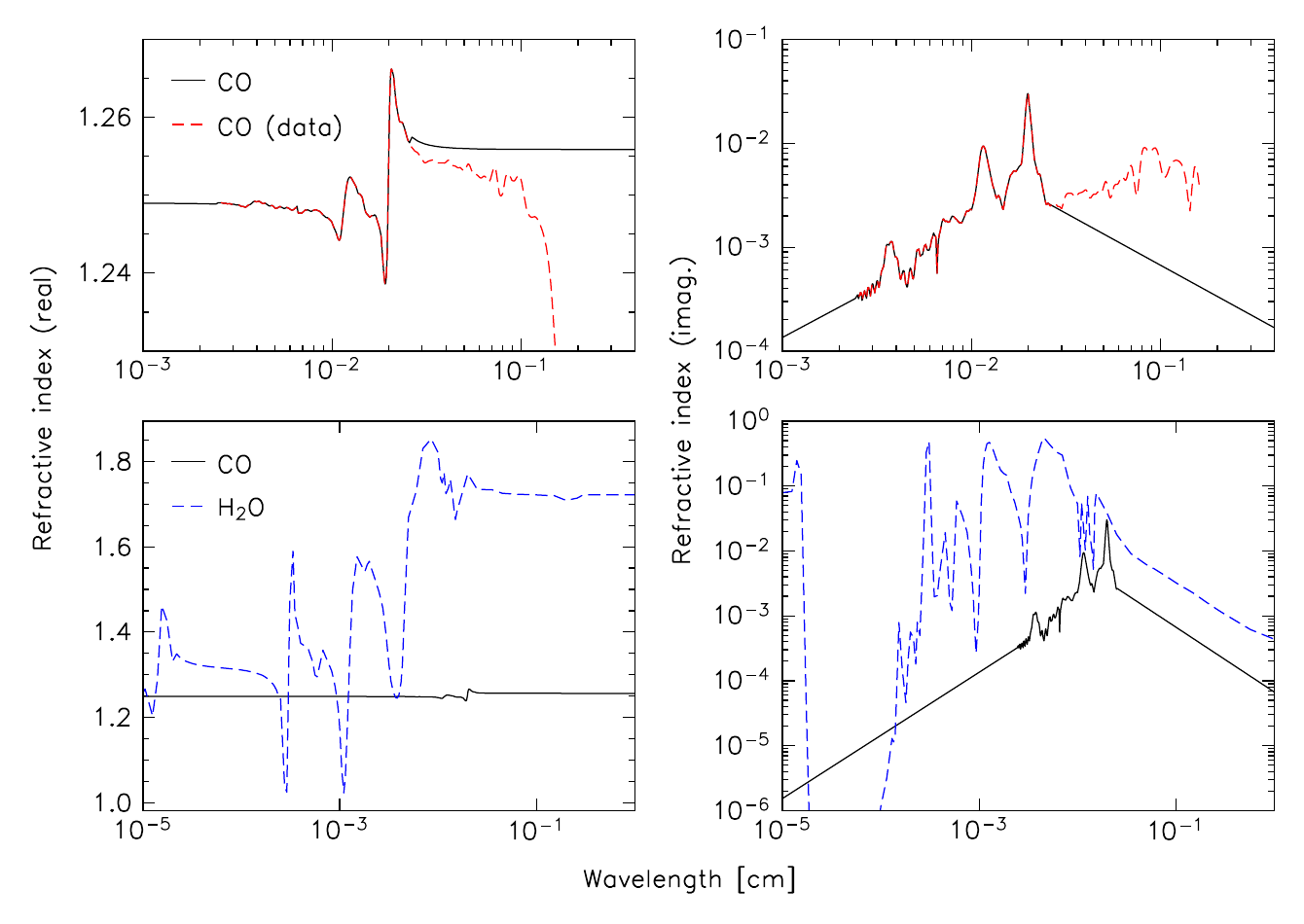}
    \caption[CO optical constants]{The optical constants used for CO when calculating opacities. The data (red dashed line) is taken from \cite{gavdush2022} and the black line shows the extended optical constants used for calculations. The optical constants for H$_2$O ice from \cite{warren2008} are plotted for comparison.}
    \label{COoptconst}
\end{figure*}

The optical constants for CO ice are adapted from the experimental data presented by \cite{gavdush2022}. The real and imaginary components of the data-derived complex refractive index, $N=n+ik$, can be seen in the red dashed lines in the upper row of Fig.~\ref{COoptconst}; for reference, the real component determines the phase velocity of waves entering the medium, i.e. scattering, whilst the imaginary component is responsible for absorption of wave energy in the medium \cite[see][]{bohren&huffman1983}. The peaks in the imaginary component can be modelled as resonant dipole excitations using Lorentz kernels; there are three absorption peaks at wavelengths of 37, 115, and 200~\unit{\micron}. The data extends from $\sim25$~\unit{\micron} to $0.2$~cm, whilst the wavelength range used in our simulations is typically 0.1~\unit{\micron} to 1~cm. To extend to the broader wavelength range, we first extend the imaginary component to short and long wavelengths, then assume that the real and imaginary components are not independent and can be related using the Kramers-Kronig relations (applicable for media that can be described by a well-defined complex refractive index):

\begin{equation}
    \label{kramersk}
    n = 1+ \frac{2}{\pi}P \int_0^\infty\frac{\nu'k}{\nu'^2-\nu^2} \text{d}\nu',
\end{equation}
where $P$ is the Cauchy principal value that allows the integral to be defined by avoiding the singularity at $\nu'=\nu$. To extend to shorter wavelengths, we perform a linear log-space extrapolation of the non-resonance background absorption below the peak at 37~\unit{\micron}. For longer wavelengths, we have chosen to cut the data at $\sim300$~\unit{\micron} and to use the long-wavelength behaviour of the Lorentz resonance kernel, i.e. 1/$\lambda$. We have chosen this method of extension for three reasons: 1) the authors state that diffraction distortions become significant at wavelengths $\lesssim0.06$~cm, 2) extrapolating from the data past the resonance peak at 200~\unit{\micron} leads to discontinuities and unphysical asymptotic behaviour of the real component, and 3) extending from the resonance peak using $k\propto1/\lambda$ is physically motivated from the Lorentz oscillator model \cite[for more discussion, see chapter 9 of][]{bohren&huffman1983}. The wavelength-extended optical constants used in simulations are shown alongside the data in the upper row of Fig.~\ref{COoptconst}, whilst the lower row shows the constants compared to those of water ice for reference.

\begin{figure*}
    \centering
    \includegraphics[width=.98\textwidth,trim={0.5cm 0.45cm .5cm 0.45cm}, clip]{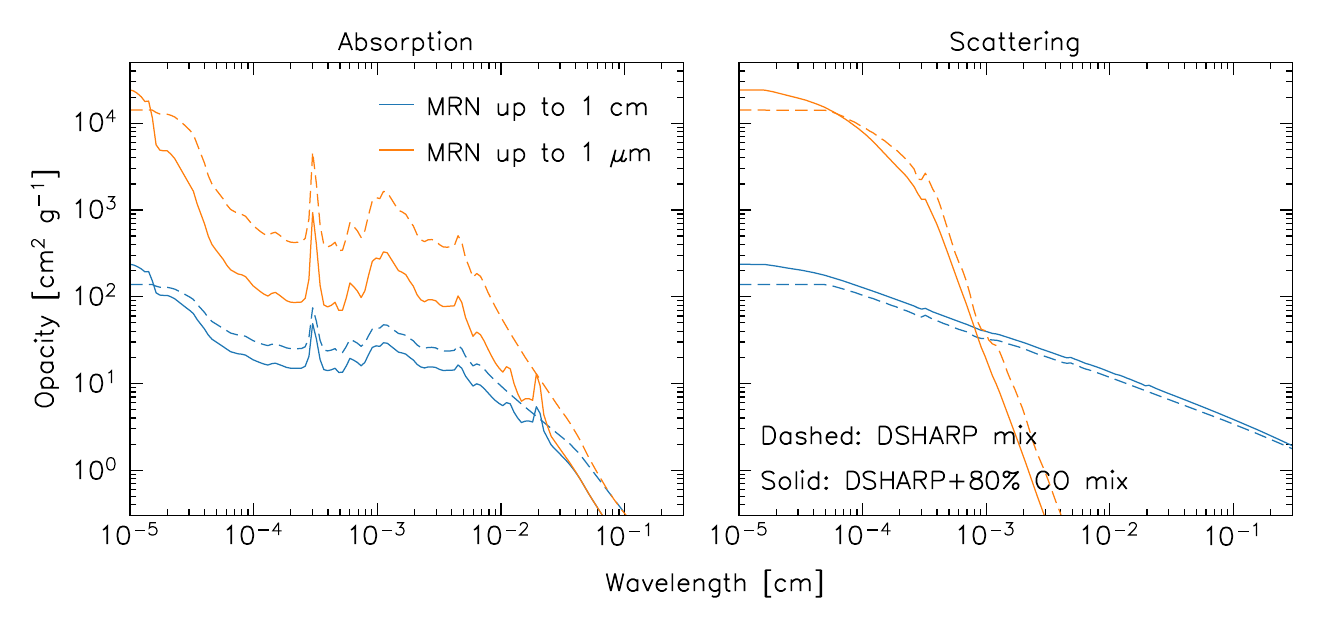}
    \caption[CO opacities]{Grain-size-distribution-averaged absorption and scattering opacities for grains with and without CO as a part of their composition. The grain-size distributions are set as MRN \citep{MRN77} distributions with maximum grain-sizes of 1~\unit{\micron} (orange lines) and 1~cm (blue lines). The base grain composition (without CO, shown by dashed lines) is the DSHARP composition \citep{DSHARP2018} and the mixed composition (solid lines) is 20\% DSHARP and 80\% CO by mass.}
    \label{COkappagrainav}
\end{figure*}

Given a set of species with optical constants and mass-fractions, we calculate an effective dielectric function, $\epsilon=\epsilon'+i\epsilon''$, using Maxwell-Garnett effective medium theory (EMT) \citep{maxwellgarnett1904}. In this model, we assume that within the heterogeneous medium, different species manifest as spherical inclusions within a homogeneous matrix. For simplifying purposes, the homogeneous matrix is assumed to have $\epsilon=1$. The composition-averaged optical constants are then used to calculate absorption and scattering opacities via 

\begin{equation}
    \label{kappa_calc}
    \kappa_{a/s} = \frac{3 Q_{a/s}}{4\rho_m a},
\end{equation}
where the absorption and scattering efficiencies, $Q_a$ and $Q_s$, are given by 
\begin{equation}
    \label{Q_a}
    Q_{a} = 12 x \frac{\epsilon'}{(\epsilon'+2)^2+\epsilon''^2},
\end{equation}
and
\begin{equation}
    \label{Q_s}
    Q_{s} = 
    \begin{cases}
        \frac{8}{3}x^4\dfrac{(\epsilon'-1)^2+\epsilon''^2}{(\epsilon'+2)^2+\epsilon''^2} & \text{for}~x< 1.3 \\
        2x^2(n-1)^2\left[1-\left(\dfrac{k}{n-1}\right)^2\right] & \text{otherwise},
    \end{cases}
\end{equation}
where $x=2\pi a/\lambda$. The two forms for $Q_s$ account for the transition between the Rayleigh regime ($Q_s\propto x^4$) for particles smaller than the wavelength, i.e. $x<1$, to the Rayleigh-Gans regime ($Q_s\propto x^2$), applicable for particles with $x\sim1$ and refractive indices of around unity (see \citealt{vandehulst}, pp. 179-182) - a reasonable assumption for the mixed-composition grains we deal with in this work. To account for anisotropic scattering, the scattering efficiency is modified by $(1-g_\text{asym})$ where $g_\text{asym}$ is the scattering asymmetry parameter. The efficiencies are limited to a maximum value of unity - this corresponds to the geometric optics regime when the absorption/scattering cross section is equal to the physical cross section. One limitation of this model is that it does not capture resonant absorption around $2\pi a/\lambda\sim 1$ for solid particles as well as Mie theory; however, this is not an important deviation for the poly-disperse grain-size distributions studied in this work; such distributions smooth over the oscillatory fluctuations seen in absorption and scattering properties of mono-disperse particles \citep{hansen1974}. In addition, as we are computing temperatures, not spectra, there is additional smoothing generated through integration over the wavelength bands. 

Fig.~\ref{cuzzimiecomp} in the Appendix shows a comparison of our opacity model to full Mie theory, calculated using the DSHARP opacity tool \citep{DSHARP2018}. There are generally only small deviations across the majority of the wavelength range that we use, making the model suitable for our thermal calculations. Fig.~\ref{COkappagrainav} shows absorption and scattering opacities calculated for grains with the base DSHARP composition and those with 80\% mass in CO ice + 20\% DSHARP. Opacities are averaged over two different MRN dust distributions, both with minimum grain sizes of 0.1~\unit{\micron}, one of ISM-like dust with a maximum grain size of 1~\unit{\micron} and the other evolved dust with a maximum grain size of 1~cm. Total opacity decreases with the addition of CO due to it having both fewer and lower amplitude absorption resonances. The effect of the most prominent CO absorption resonances (at 115 and 200~\unit{\micron}) can be seen in the absorption opacity.

\section{The snow line instability}\label{snow linesinstab}

The snow line instability is similar in nature to the well-known thermal instability of accretion discs \cite[see][]{shakura1976, pringle1981}. The simple picture of the thermal instability can be understood by considering the relative changes of the heating and cooling rates in the disc at a certain radius, $q_+$ and $q_-$, with respect to small changes in the disc's temperature at that location. If the cooling rate increases with temperature by a greater amount than the heating rate, the system will return to equilibrium for any perturbation to the temperature. However, if the heating rate increases with temperature by a greater amount than the cooling rate, a small increase (decrease) in temperature will lead to runaway heating (cooling). The criterion for a location in the disc to be unstable can be written as \citep{pringle1981}

\begin{equation}
    \label{instabcriterion}
    \frac{\text{d} \log q_+}{\text{d} \log T_m} > \frac{\text{d} \log q_-}{\text{d} \log T_m},
\end{equation}
where $T_m$ is the mid-plane temperature of the disc. 

In the two-layer approximation for passive heating, the heating and cooling rates can be written as 
\begin{equation}
    q_+ = \frac{1}{2}\psi_s F_*, \hspace{1cm} q_-=\psi_i\sigma T_m^4,
\end{equation}
where $F_*$ is the intercepted stellar flux, given by
\begin{equation}
    \label{F_*}
    F_*  = \dfrac{L_*}{4 \pi R^2} \sin \alpha_f,
\end{equation}
and $\psi_{s/i}$ account for the fraction of radiation from the surface/interior that is absorbed or emitted by the interior \citep{dullemond2001}. 

If the disc interior is optically thick to the given radiation, then $\psi_{s/i}$ are well approximated as unity. \hl{When the disc is optically thick to both surface and interior radiation, the disc is stable. This is because the cooling rate scales as $T_m^4$, while $F_*$ is only weakly dependent on $T_m$ via the flaring angle, $\alpha_f$.}

When the interior is optically thin, the absorption fraction can be written as $\psi\approx\tau \approx \Sigma_\text{sol}\kappa$, $\Sigma_\text{sol}$ and $\kappa$ being the total solid surface density and mean opacity for the appropriate radiation field. We neglect any gaseous contribution because the gas opacity is negligible in comparison to the solids. If the interior is now optically thin to both surface and interior radiation, we are still left with a stable disc because both heating and cooling are proportional to $\Sigma_\text{sol}\kappa$. However, if the disc is optically thick to surface radiation but optically thin to interior radiation, then the cooling rate is proportional to $\Sigma_\text{sol}\kappa T_m^4$ whilst the heating rate isn't. Under conditions where $\Sigma_\text{sol}\kappa$ decreases sufficiently steeply with increasing $T_m$, we can overcome the stabilisation generated by $T_m^4$. 

\hl{Discs may therefore become thermally unstable at snow lines, where a small increase in temperature causes a large amount of volatiles to evaporate, substantially reducing the optical depth and cooling rate.} If we assume that the opacity remains constant (we know already that this is not the case, but we can assume that changes in total solid density are more important), then we can now write the condition for instability, Eqn.~\ref{instabcriterion}, as

\begin{equation}
    \label{instabcriterion2}
    \frac{\text{d} \log \Sigma_\text{sol}}{\text{d} \log T_m} \lesssim -4.
\end{equation}


\hl{Sublimation of ices can readily meet this requirement. For example, consider we have 50\% of our solid mass in CO ice at the snow line, and we perturb the temperature up by 15\%, say from 20 to 23~K; in doing so, we cross the sublimation temperature and all of the CO sublimates.} In this case, the left-hand side of Eqn.~\ref{instabcriterion2} is $-5$ and the disc is unstable. \hl{If the amount of solids near the snow line decreases, then the cooling rate decreases causing the temperature to rise, further decreasing the solid density and producing a phase of runaway heating.} This runaway heating would manifest in the disc as the snow line rapidly moving to larger radii as the ice sublimates. Ultimately, the snow line will stop moving once it reaches a radius at which the disc becomes stable; this can either be where the opacity conditions described above break down, 
or at the point where all of the ice has sublimated and $\frac{\text{d} \log \Sigma_\text{sol}}{\text{d} \log T_m} \approx 0$. 
The opposite will happen for a decreasing temperature perturbation, with the snow line moving radially inwards until it reaches a region where the disc is too warm for any ice to remain, and the disc is again stable. 
\hl{These boundaries therefore define the edges of the unstable region, within which there is a range of radii for which there are two solutions for the temperature}; one with a ``hot'' equilibrium temperature and all of the volatiles in the vapour phase, and one with a ``cold'' equilibrium temperature and all of the volatiles in the ice phase. 

\hl{These two temperature solutions are associated with different mass-fluxes of volatiles due to differences between the radial velocities of the vapour and ice phase volatiles.} When in the ice phase, the volatiles are coupled to the dynamics of the dust population --- which radially drifts relative to the gas --- and we therefore expect the radial velocity to be larger than that of the vapour-phase volatiles. A difference in radial velocity means a difference in radial mass-flux between the two solution branches; the hot, vapour-phase solution being a low mass-flux state, whilst the cold, ice-phase solution is a high mass-flux state. 

As discussed in \cite{owen2020}, this \hl{separation} in mass flux can allow a limit-cycle to be established, where the disc is driven to jump between the hot and cold states in a periodic fashion. Consider a radius in the unstable region that is in the cold, high mass-flux state and is being fed by a lower ice mass-flux from the outer disc. Over time the volatile surface density at that radius will decrease, and the decrease in ice will lead to an increase in temperature due to the opacity configuration. At a critical ice density, the disc will be hot enough to trigger sublimation of the ice into the vapour  phase and therefore a jump to the hot, low mass-flux state. The disc at this radius is still being fed by the same ice mass-flux from the outer disc, but this now leads to a build-up of volatiles over time, and therefore a decreasing temperature. As before, we reach a critical point where the disc is cool enough that all of the vapour can condense into ice, and we jump back to the cold, high mass-flux state.

In simulations performed by \cite{owen2020}, this limit-cycle manifests as the CO snow line moving through 10s of au on kyr timescales as local regions in the disc transition between the hot and cold states. They studied the snow line using 1D radial models, where evolution of the dust grain-size distribution was modelled by solving the time-evolution of the maximum grain-size, motivated by \cite{birnstiel2012}. One limitation to this approach is that it ignores any impact that vertical fluxes of CO may have on setting up the instability and allowing a limit-cycle to exist, as it does not take into account the 2D structure of the snow surface generated by the vertical temperature structure. In addition, temperature perturbations at the mid-plane affect the vertical structure and therefore could lead to shadowing of the outer disc and potentially the spawning of more features. We also expect the increased complexity to which we model the evolution of the grain-size distribution and the deposition of ice to impact the results, as we have already seen in section~\ref{icevapourchemistry} that the distribution of ice across the grain distribution can vary depending on the growth timescale. With our model, we are now able to investigate the impact that both of these effects -- vertical structure and detailed grain-size distribution evolution -- have on the results found previously.

\subsection{Static equilibrium solutions in 2D}

To investigate the theoretical instability in discs with a resolved vertical dimension, we set up a disc system with only 10~\unit{\micron} dust grains and, starting from either a cold or hot disc, iterate the temperature and vertical gas and dust structures towards equilibrium solutions. The grid parameters are: 600 power-law-spaced radial cells ($R^n$, $n=0.25$) between $R=0.3$ and $300$~au, and 250 power-law-spaced cells ($\theta^n$, $n=0.75$) between $\theta=0$ and $\pi/6$~rad. The stellar properties are shown in Table~\ref{tab:stellarprops}. Stellar heating calculations use 200 logarithmically-spaced wavelength bins, spanning a wavelength range of $0.1-10^5$~\unit{\micron}. These wavelengths, and the opacities associated with them, are binned into 40 logarithmically-spaced wavelength bands for the flux-limited-diffusion (FLD) calculations used for radiative transport of re-emitted thermal radiation. A constant background temperature of 10~K is applied to provide a baseline radiation field that bathes the entire disc. We set the relative tolerance of the FLD solver to $10^{-5}$. The gas is initialised with the \cite{lbp74} self-similar solution with a characteristic radius of 50 au and an initial gas mass of 0.06 M$_\odot$. The general disc parameters are: $\alpha=\,$\tenpow{-3}, dust-to-gas ratio = \num{6e-4}. This low dust-to-gas ratio is necessary because the opacity of 10~\unit{\micron} grains is much larger than that of large grains (cf. Fig.~\ref{COkappagrainav}), \hl{and we need optical depths of <~1 to the ~60 K radiation field emanating from the surface to fulfil the moderate optical depth requirement of the instability discussed in section 3}. The CO abundance relative to atomic hydrogen, CO/H, is set to \num{8e-5}, just under the canonical value; this leads to an ice mass-fraction of $\Sigma_{\text{i,CO}}/\Sigma_{d+\text{i,CO}}=0.75$, where $\Sigma_{\text{i,CO}}$ and $\Sigma_{d+\text{i,CO}}$ are the surface densities of CO ice and total solids (dust and ice) respectively. This ice mass-fraction is high compared to other studies (e.g. Öberg et al. 2021); this is because we approximately maintain the ISM CO abundance whilst reducing the dust-to-gas ratio. \hl{For this static model, where we simply want to demonstrate that two solutions can exist, a high ice mass-fraction is justified to enhance the effect of the ice on the opacity structure --- in the dynamic models, we use a more standard ice mass-fraction of 20\%. Regardless, the chosen ice mass-fraction does not affect the existence of the limit-cycle; this is discussed in section 4.3.}

\begin{table}
    \centering
    \begin{tabular}{ccc}
        \hline Stellar property &  Value \T \B \\ \hline
        \T Mass, $M_\star$ & 1~M$_\odot$ \\ 
        Effective temperature, $T_\star$ &  4500~K \\
        Radius, $R_\star$ & 1.7 R$_\odot$ \\ \hline
    \end{tabular}
    \caption{Stellar properties for both the static and dynamic models. We assume a solar-mass pre-main-sequence star for all of our models.}
    \label{tab:stellarprops}
\end{table}

When starting in the hot state, the initial temperature profile is set to $T=100(R/\text{au})^{-0.4}$~K, for which the snow line is at a radius of $\sim50$~au; for the cold state, $T=80(R/\text{au})^{-0.5}$~K, for which the snow line is at $\sim15$~au. As the temperature is iterated from each initial condition slowly towards equilibrium, the gas and dust vertical profiles are updated respectively through solving for hydrostatic equilibrium and solving for diffusion-settling equilibrium (see Eqn.~30 in \citealt{taklin02}). The ice-vapour chemistry is also evolved to equilibrium at each iteration. As the simulations iterate towards convergence, the snow lines move radially inwards/outwards for the initially hot/cold discs. If the disc is unstable, the two snow line solutions lie at the edges of the unstable region in the disc. When we advance the snow lines from initially hot/cold states (making sure to force the temperature structure to evolve slowly), they should therefore stop at the first equilibrium solution they encounter, thereby locating the two possible solutions.

\begin{figure}
    \centering
    \includegraphics[width=\columnwidth,trim={0.55cm .5cm 0.55cm 0.25cm},clip]{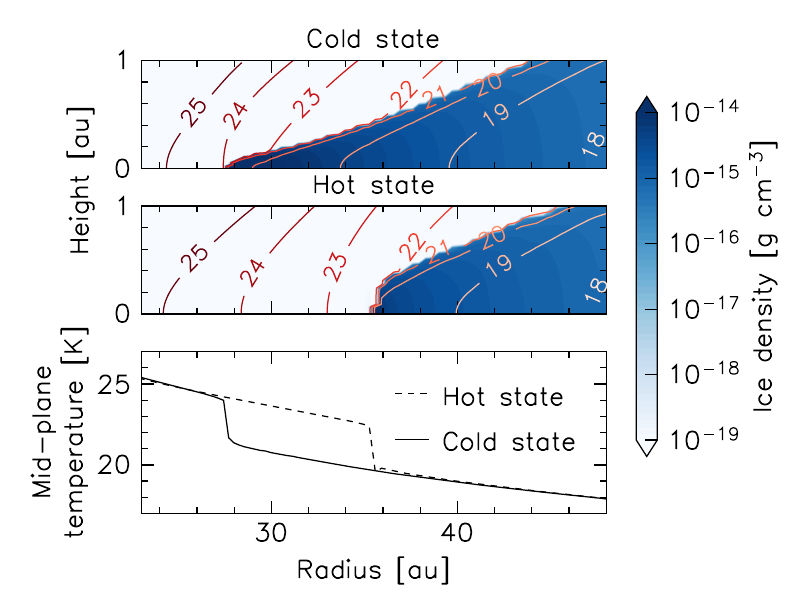}
    \caption[2D Ice density and temperature structure of single-grain equilibrium snow line solutions]{2D ice density and temperature structure of stable equilibrium snow line solutions in the single-grain size model. Temperature contours (in K) are over-plotted on the density maps.}
    \label{eqbmsols}
\end{figure}

\begin{figure}
    \centering
    \includegraphics[width=0.99\columnwidth, trim={0.5cm 0.2cm 0.4cm 0.2cm}, clip]{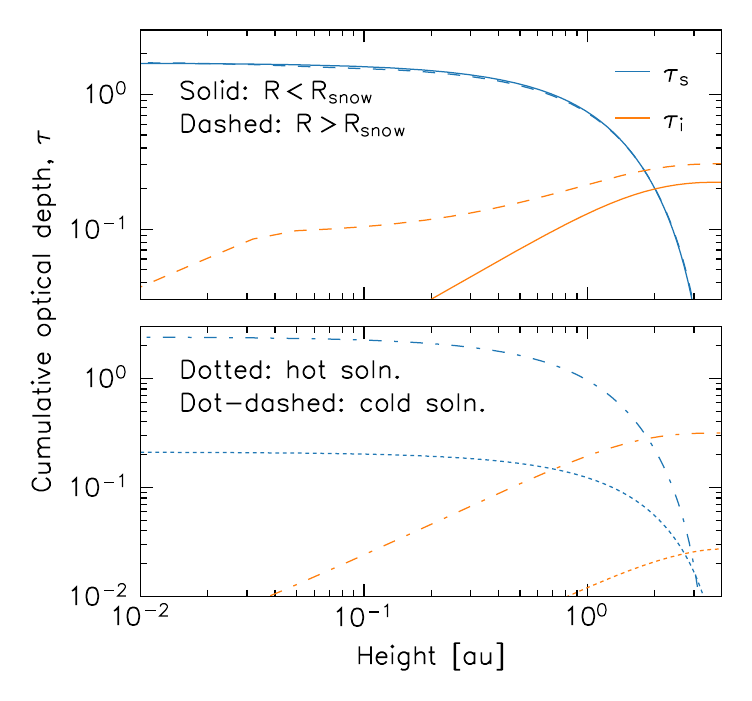}
    \caption[Cumulative optical depths around snow lines]{Cumulative optical depths to mid-plane and surface radiation, $\tau_i$ and $\tau_s$, as functions of height. For the optical depth to mid-plane radiation, the column is summed from the mid-plane up to the surface, and vice versa for the optical depth to the surface. The top panel shows how the optical depth changes on either side of the snow line, which lies at radius $R_\text{snow}$, due to ice condensation affecting the solid mass. The bottom panel shows the optical depths at the snow lines when the system has converged to a single equilibrium solution, either hot or cold.}
    \label{COopticaldepth}
\end{figure}
Fig~\ref{eqbmsols} shows two stable equilibrium solutions found through this method, separated by $\sim8$~au. The drop in mid-plane temperature at the snow line can be understood by examining the changes in optical depth caused by the increase in solids at the snow line. Calculating the optical depth of the disc interior to both the surface and the mid-plane radiation allows us to investigate how well the disc at altitudes above the mid-plane absorbs the mid-plane radiation field and how well the disc interior absorbs the surface radiation field. To calculate the optical depth, we evaluate the Planck mean opacity of the disc at a given temperature (surface/mid-plane); this is appropriate to use given we are in a marginally optically thin regime. The top panel of Fig.~\ref{COopticaldepth} shows the cumulative optical depth in the cells either side of the radial location of the mid-plane snow line as functions of height and at two different radiation temperatures: that of the mid-plane, at 22~K; and of the surface, at 60~K. These values are taken from the vertical temperature structure at the snow line. The cumulative optical depth at 22~K is found by summing upwards from the mid-plane to the surface, and vice versa for the depth at 60~K. The optical depth to the surface/interior radiation, $\tau_{s/i}$, in a cell is calculated via

\begin{equation}
    \label{opticaldepth}
    \tau_{s/i} = (\rho_\text{d}+\rho_\text{i})\kappa_{P,s/i} \Delta Z,
\end{equation}
where $\kappa_P$ is the Planck mean opacity, given by
\begin{equation}
    \label{kappa_planck}
    \kappa_{P,s/i}=\frac{\int\kappa_{a,\nu} B_\nu(T_{s/i}) \text{d}\nu}{\int B_\nu(T_{s/i}) \text{d}\nu},
\end{equation}
where $\kappa_{a,\nu}$ is the frequency-dependent absorption opacity for the given grain composition. As described in section~\ref{snow linesinstab}, the instability requires the disc interior to at least be marginally more optically thick to the surface radiation than the disc above the mid-plane is to the mid-plane radiation. The top panel of Fig.~\ref{COopticaldepth} shows that at radii just interior to the snow line (i.e. $R<R_\text{snow}$), the disc mid-plane is optically thick to the surface radiation, with $\tau_s>1$, and the disc above the mid-plane is marginally optically thin to the mid-plane radiation, with $\tau_i\sim0.2$. As we move to the next radial cell, exterior to the mid-plane snow line (i.e. $R>R_\text{snow}$), we see that the increase in solid density around the mid-plane due to ice causes an increase in $\tau_i$. Given that the emissivity of an optically thin medium is proportional to the optical depth, this increase in depth increases the cooling rate relative to that of heating (which is optically thick), causing the disc to equilibrate at a cooler temperature. 

From this system of two stable equilibria for a given volatile surface density, we can modulate the surface density to explore the density range over which the disc is unstable. As the density is increased, $\tau_i$ will increase, and at some point be large enough that the cooling rate will no longer respond more sensitively to temperature; at this point we would expect the disc to be globally stable with only one snow line solution. Likewise, decreasing the density will decrease $\tau_s$; once this is also optically thin enough that the heating rate responds as sensitively to temperature as the cooling, we again expect to see a single snow line solution. 

\begin{figure}
    \centering
    \includegraphics[width=\columnwidth,trim={0.2cm .2cm 0.2cm 0.2cm},clip]{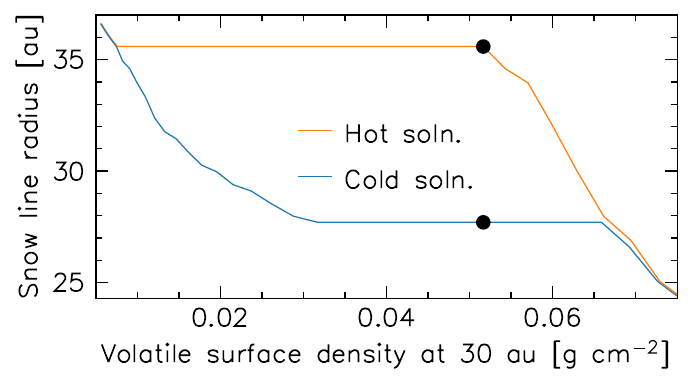}
    \caption{Mid-plane snow line radii for the hot and cold solutions as a function of total volatile surface density. We pick a reference radius of 30~au for the total volatile surface density as this lies between the snow line solutions. The black circles indicate the two snow line solutions for the initial surface density.}
    \label{hc_conv}
\end{figure}

When we increase/decrease the total surface density of solids (ice + dust) in the simulations, the cold/hot solution remains in place until the solutions converge and begin to move in tandem, as can be seen in Fig.~\ref{hc_conv}. The bottom panel of Fig.~\ref{COopticaldepth} shows the optical depths at the snow lines (in the final cell where $R<R_\text{snow}$; therefore there is no contribution to the optical depth from the ice) for the systems where the two solutions have converged to one through decreasing and increasing the total solid surface density. The dotted lines show the single solution found when the surface density is decreased, i.e. the disc is in the hot state, whilst the dot-dashed lines show the single cold state solution for an increased surface density. The solutions converge to the hot equilibrium when the optical depth to the surface radiation drops to $\sim0.2$, whilst the solutions converge to the cold solution when the optical depth to mid-plane radiation surpasses $\sim0.3$. This implies that, for this setup, when $\tau_s,\tau_i \lesssim 0.2$, they are both optically thin enough for the heating rate to respond more sensitively to temperature than the cooling rate, and vice versa for $\tau_s,\tau_i \gtrsim 0.3$ - i.e. the transition between optically thick and thin occurs in the optical depth range of $0.2-0.3$.

\section{Dynamical evolution}

We now construct simulations to study whether the CO snow line can be unstable in a 2D dynamical disc, and whether a limit-cycle can be established. 

\subsection{Methods}

These simulations have a similar set up to the equilibrium calculations detailed in the previous section, but with some notable differences. We use 300 power-law-spaced radial cells ($R^n$, $n=0.25$) between $R=0.5$ and $100$~au, and 180 power-law-spaced cells ($\theta^n$, $n=0.75$) between $\theta=0$ and $\pi/6$~rad. A full distribution of dust grains is included, with a resolution of 7 mass-bins per decade, to give 148 bins logarithmically-spaced between 0.1~\unit{\micron} and 100~cm. To increase the computational efficiency, we run \hl{radial} 1D gas and dust models in the inner region of the grid, from 0.5 to 15~au, which couple to 2D calculations in the remaining outer disc. Using a 1D inner region significantly speeds up the simulations; both by reducing the dimensionality of the problem in the 1D region, but also because the 2D CFL time-step is typically governed by diffusion in the vertical dimension in the mid-plane cell at the smallest radius, whilst the 1D CFL time-step is governed by diffusion in the radial dimension, which is much less constraining. The 1D solver updates are sub-cycled within the global 2D CFL time-step loop in the same fashion as the coagulation and temperature updates \cite[see][]{robinson2024}. This approach is valid for studying the CO snow line, as it resides at radii $\gtrsim20$~au, and therefore the detailed 2D dynamics in the inner disc are not important for this problem; we simply require it for calculating the correct temperature profile. For details of the 1D methodology, see Appendix~\ref{1Dmethods}. 

To facilitate the 1D-2D grid interface, we have implemented a sub-gridding method to the code wherein radial portions of the full grid can be separated into sub-grids. Making the sub-grids portions of the full grid allows for simple copying of quantities between the full grid and the sub-grids, and makes the ghost cells at the boundaries of the sub-grids overlap identically with active cells on the full grid, making boundary conditions easier to set. Sub-grids can be either 1D or 2D; if 1D, the sub-grid shares a portion of the radial grid of the full 2D grid and only has one active cell in the vertical dimension.

The viscosity is set to be linear with respect to radius, i.e. $\nu=\nu_0(R/\text{au})$. As described in \cite{robinson2025}, a linear relationship with radius is fairly consistent with observations \citep[see e.g.][]{hartmann1998} and the underlying nature of the viscosity is not well constrained --- therefore coupling the viscosity to the evolving temperature structure is unmotivated. $\nu_0$ is calculated using the $\alpha$ formalism \citep{shakira1973}, $\nu_0=\alpha c_{s,\text{au}}^2/\Omega_{K,\text{au}}$~\unit{\cm\squared\per\second}, where $c_{s,\text{au}}$ is the sound speed at 1 au assuming a mid-plane temperature of 160~K at 1 au, and $\Omega_{K,\text{au}}$ is the Keplerian angular velocity at 1 au.

\subsubsection{Boundary conditions}

\begin{figure*}
    \centering
    \includegraphics[width=0.99\textwidth, trim={0.25cm 0.1cm 0.1cm 0.1cm}, clip]{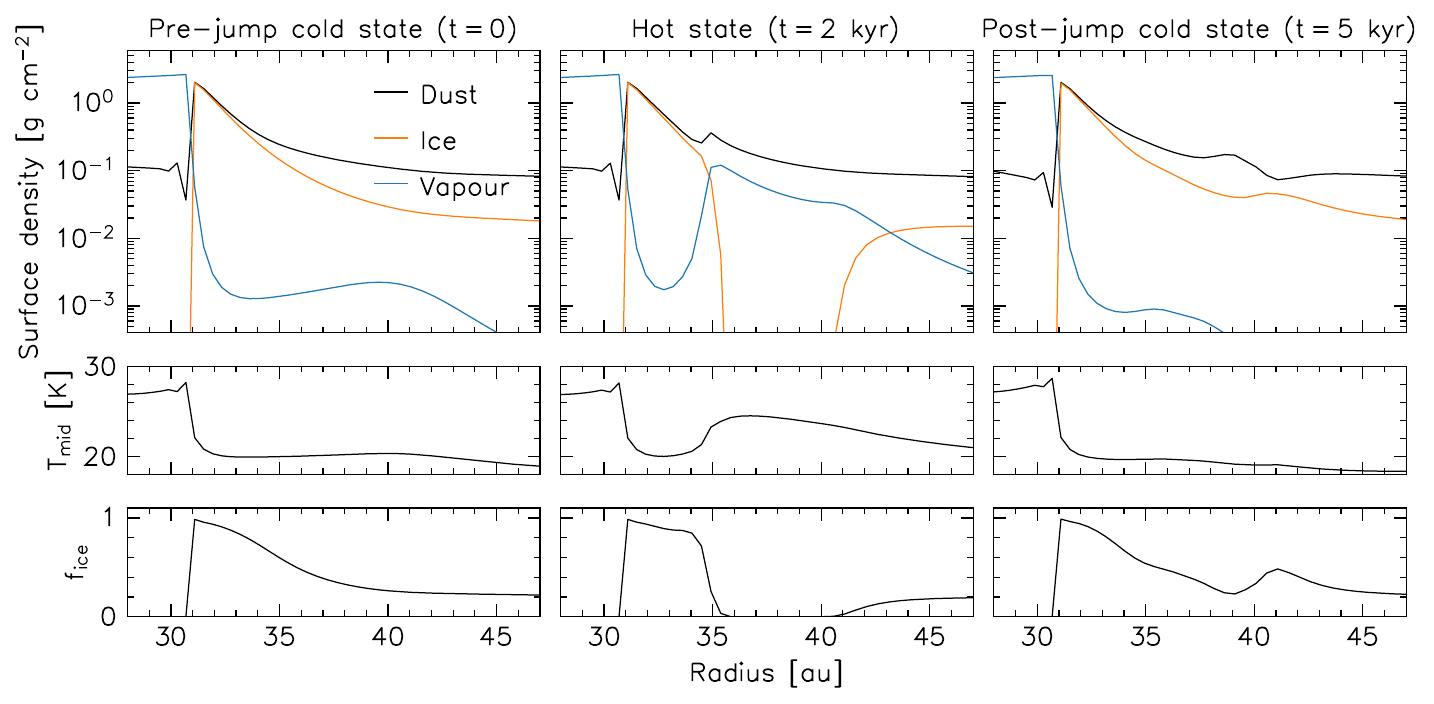}
    \caption[Surface density evolution]{Snapshots of the vertically-isothermal simulation in the pre-jump cold state, during the hot state, and in the post-jump cold state. The top row shows the surface densities of dust (ice included), ice and vapour. The middle row shows the mid-plane temperature, $T_\text{mid}$, whilst the bottom row shows the ice mass-fraction of the dust, $f_\text{ice}$.}
    \label{sigmaevo_isoT}
\end{figure*}

The outer boundary condition on the dust surface density in the 1D grid is set by the vertically-integrated dust density from the 2D grid, i.e. 

\begin{equation}
    \label{Sig1DBC}
        \Sigma_\text{1D}^\text{GC} = \int\rho_{\text{2D}}^\text{AC} \text{d}Z,
\end{equation}
where GC and AC stand for ghost cell and active cell, as the ghost cells of the inner 1D grid overlap with the first active cells of the 2D grid. The inner boundary condition on the dust density in the 2D grid is set using the surface density from the 1D grid projected into the vertical dimension using the normalised vertical dust distribution in the first active cell of the 2D grid, i.e.

\begin{equation}
    \label{rhoBC}
    \rho_\text{2D}^\text{GC} = \Sigma_\text{1D}^\text{AC} \left(\frac{\rho_\text{2D}^{\text{AC},0}}{\Sigma_\text{2D}^{\text{AC},0} }\right),
\end{equation}
where AC$,0$ corresponds to the first active cell of the 2D grid. The 1D and 2D velocities are then set via standard outflow boundary conditions. In contrast to the default choice for the other boundaries, diffusion over the 1D-2D grid boundary is allowed. Expanding the inner 1D densities to 2D for calculating the temperature structure is done in the same fashion as the static equilibrium calculations; the gas through solving for hydrostatic equilibrium and the dust through solving for diffusion-settling equilibrium. The 2D densities at the interface between inner and outer grids are joined together through a smoothing function taken from \cite{paxton2011}; defining $F=\log(R/R_\text{in})/\log(R_\text{out}/R_\text{in})$ and $S=[1-\cos(\pi F)]/2$, the smoothed density is calculated via

\begin{equation}
    \label{smoothrho}
    \log\rho_\text{smooth} = S\log\rho_\text{outer} + (1-S)\log\rho_\text{inner},
\end{equation}
where $\rho_\text{inner}$ is the surface density in the outer, 2D region expanded in the vertical dimension using the diffusion-settling equilibrium calculation used in the inner region, whilst $\rho_\text{outer}$ is the actual 2D density profile of the outer region. For this model, $R_\text{in}=15$~au and $R_\text{out}=17$~au. The dust distribution is also smoothed in the inner regions of the 1D grid to avoid a pile-up of dust that leads to an unstable temperature profile due to shadowing. Using the same smoothing formalism defined above, the dust densities are smoothed down to a dust-to-gas ratio of \tenpow{-6} at 0.5~au, with $R_\text{out}=1.2$~au. The dust distribution is initiated as an MRN distribution with a cut-off in grain size at the minimum of the fragmentation and drift limits. 

To study whether the disc enters a limit-cycle, we want the mass-fluxes of dust and volatiles at the outer boundary to be constant with time so that any dynamical evolution is driven by the instability rather than global evolution of the dust and volatile distributions. In a real disc, this would not be the case due to the evolution of the dust distribution, the conversion of dust to planetesimals, and the accretion of solids onto the central star. However, we expect that there should be phases of quasi-steady volatile mass-flux that exist for timescales greater than those associated with the limit-cycles (kyr), meaning they would still have an impact on the disc structure even if they did not exist for the entire lifetime of the disc. To implement fixed mass-fluxes of dust and volatiles at the outer boundary, the dust and volatile densities and velocities in the outer 5~au are kept constant at the initial values. The gas is also assumed to be in steady-state and therefore kept at the initial values of surface density; the volume density and velocity structure are however updated as the temperature structure evolves so as to maintain a constant gas mass accretion rate over time.

\subsection{Vertically-isothermal ice-vapour solutions}\label{vertisorun}

To compare to the 1D simulations of \cite{owen2020}, we first run calculations where the ice-vapour solver is fed a vertically-isothermal temperature structure; i.e. the 2D temperature field that is fed to the ice-vapour solver at each update is a copy of the full 2D temperature field where the temperature at all altitudes is set to the mid-plane value. Doing this means that the 2D structure of the snow surface is controlled purely by the vapour density profile, and there is therefore negligible vapour present at altitudes above the snow surface at radii exterior to the mid-plane snow line radius. \hl{Whilst the temperature structure is comparable to that in \cite{owen2020} in being 1D, the simulations do differ by our inclusion of the full grain-size distribution.}

In such a setup, we see that the disc does enter a limit-cycle --- albeit one that is phenomenologically different to that found by \cite{owen2020}. Our fiducial simulation uses parameters very similar to those in the fiducial simulation of \cite{owen2020}: $\alpha=10^{-3}$, $v_\text{frag}=10~$\unit{\metre\per\second}, $M_\text{disc}=0.05~$M$_\odot$, and outer disc values of the ice mass-fraction and dust-to-gas ratio of 20\% and 1\% respectively. Fig.~\ref{sigmaevo_isoT} shows how the surface densities of each species, the mid-plane temperature and the ice mass-fraction evolve throughout the cycle. The disc remains in the cold state for $\sim100~$kyr, before jumping to the hot state for $\sim3$~kyr. The hot state is different to that seen by \cite{owen2020}, as the disc only jumps to the hot state some distance behind the mid-plane snow line, leading to two isolated snow surfaces. The hot state is short-lived, as vapour rapidly builds up in the hot region due to the advection of icy pebbles from the outer disc and diffusion of icy dust outwards from the inner cold region. As the disc remains in the hot state for only a short time, the perturbation in the dust surface density that it triggers is diffusively smoothed out on short timescales after the disc has returned to the cold state; it therefore has no subsequent impact on the dust distribution interior to the cold state mid-plane snow line.
\begin{figure}
    \centering
    \includegraphics[width=0.99\columnwidth, trim={0.25cm 0.1cm 0.25cm 0.1cm}, clip]{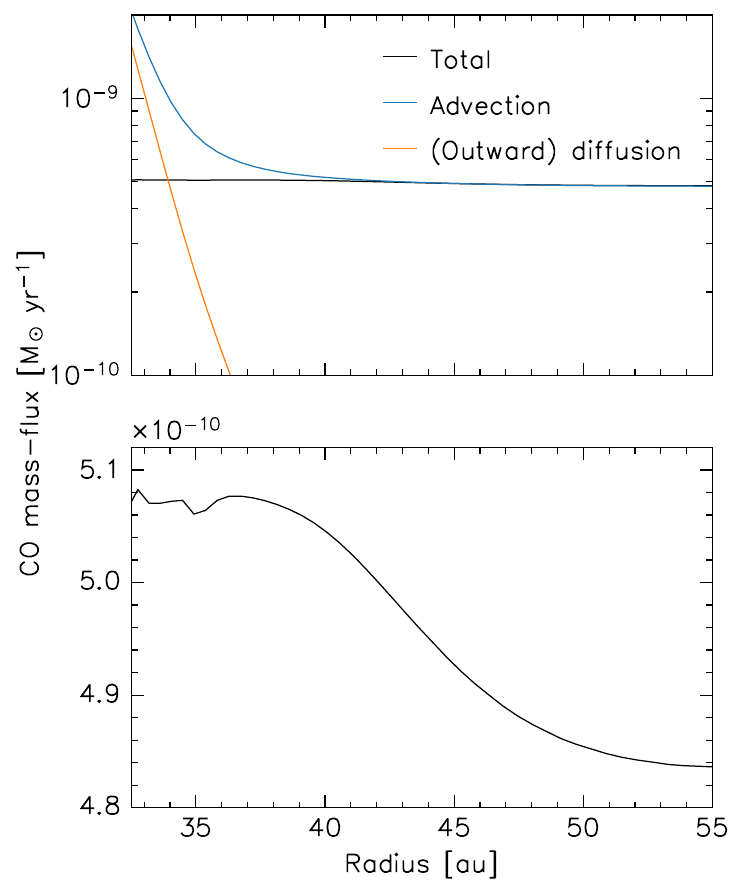}
    \caption[CO mass-flux]{The top panel shows the total CO mass-flux as a function of radius whilst the disc is in the cold state, along with the individual advective and diffusive flux components; the mass-flux is defined as positive in the radially-inward direction, i.e. towards the star, but for plotting purposes, the diffusive flux is plotted as positive in the radial-outwards direction. The bottom panel shows the same total CO mass-flux (solid line), but zoomed-in for clarity.}
    \label{massflux_isoT}
\end{figure}

\hl{As discussed in section 3 for the models presented in \cite{owen2020}, the instability is triggered by the disparity between ice and vapour mass-fluxes in the vicinity of the snow line. During the cold state, at the snow line, vapour diffuses radially outward and condenses onto small grains. Due to the long growth time, these small icy grains diffuse outwards by several au before being swept up by large grains. As the large grains rapidly drift inwards and sweep up the icy grains, they lead to a large inwards advection of ice. This can be seen in the upper panel of Fig.~\ref{massflux_isoT}, where we see a large radially outwards diffusive flux of small icy grains that is countered by a radially inwards advective flux of large icy grains. Whilst these two fluxes appear to cancel out on the scale shown in the upper panel of Fig.~\ref{massflux_isoT}, with the tighter y-axis limits used in the lower panel we see that there is structure in the net mass-flux of CO. From $\sim36$ to $\sim55$~au, the net mass-flux decreases as a function of radius, due to the advective flux of large grains. The ice mass-fraction therefore decreases as a function of time in this radial region where the net mass-flux has a negative gradient, as the advective flux of icy grains moving to smaller radii is smaller than that which is replenishing the region from the outer disc past $\sim55$~au. }
\begin{figure}
    \centering
    \includegraphics[width=0.99\columnwidth, trim={0.2cm 0.1cm 0.2cm 0.1cm}, clip]{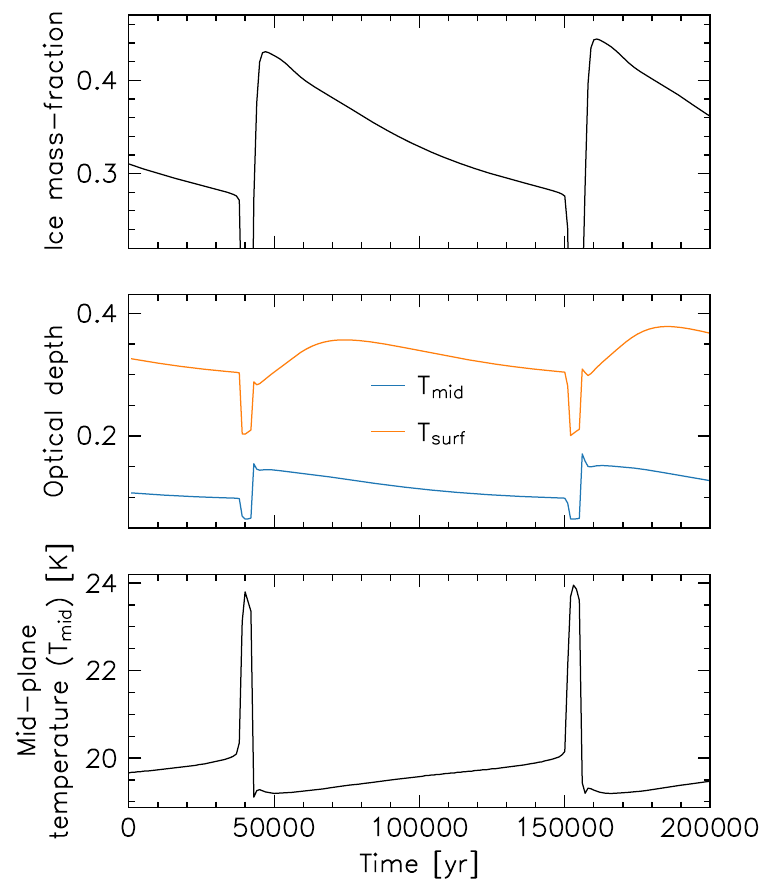}
    \caption[Optical depth]{The ice mass-fraction, vertical optical depth to surface and mid-plane radiation fields ($T_\text{surf}~\&~T_\text{mid}$ respectively), and mid-plane temperature as functions of time, at a radius of 39.6~au. After each jump back to the cold state from the hot state (at around 42~kyr and 155~kyr), the optical depth to surface radiation takes a longer amount of time to reach a maximum than that to the mid-plane radiation. This is because the freshly-deposited ice takes time to grow into the $\sim100~$\unit{\micron} grains that dominate the opacity at $T_\text{surf}$. The opacity to the mid-plane radiation is not as affected by this, as a large contribution to the opacity comes from the CO absorption peaks at $\sim100-200~$\unit{\micron} that are substantial even for very small icy grains.}
    \label{tauvst_isoT}
\end{figure}

\begin{figure*}
    \centering
    \includegraphics[width=0.99\textwidth, trim={0.1cm 0.1cm 0.2cm 0.1cm}, clip]{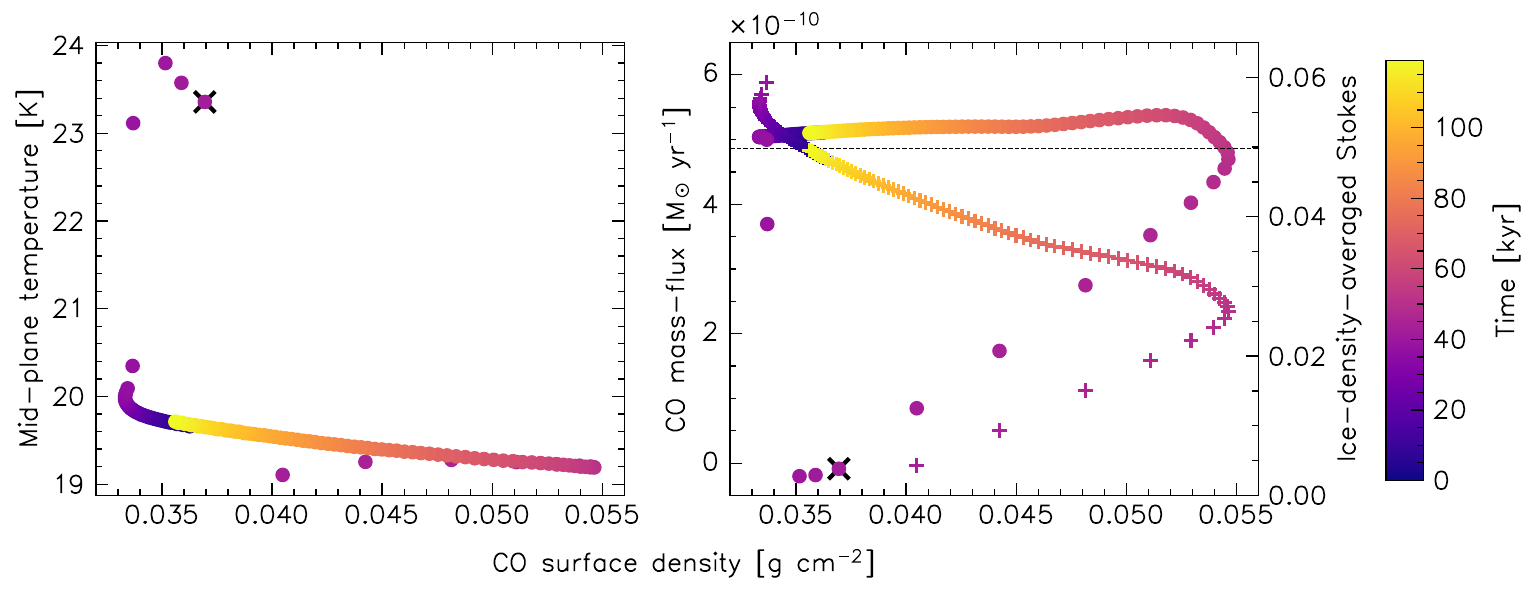}
    \caption[S-curves]{S-curves of stable equilibrium solutions \hl{for one cycle} in mid-plane temperature (left panel), CO mass-flux, (right panel circles) and ice-density-averaged Stokes number (right panel crosses) vs. CO surface density space for the vertically-isothermal simulation, at a radius of 39.6~au. There are no values for the average Stokes number whilst the disc is in the hot state. The points marked with black crosses correspond to the final snapshot where the disc is in the hot state. The dotted line in the right panel corresponds to the mass-flux from the outer disc that is feeding the unstable region.}
    \label{scurve_isoT}
\end{figure*}

\hl{The decrease in ice mass-fraction in this radial region during the cold state can be seen in Fig.~\ref{tauvst_isoT}, which plots how the ice mass-fraction, optical depth and mid-plane temperature evolve over the course of the limit-cycle at a radius of 39.6 au. As the ice mass-fraction decreases, attempting to drive the mass-flux to equilibrium, the optical depths to both surface and mid-plane radiation likewise decrease. The heating via surface radiation is more optically thick than the cooling via mid-plane radiation, meaning that, as both optical depths decrease, the efficiency of cooling decreases more than that of heating, and the mid-plane heats up. As the desorption rate is exponentially sensitive to temperature, at a certain point the ice starts rapidly sublimating, causing runaway heating; this is the jump to the hot state. When in the hot state, large icy grains drifting in from the outer disc and small icy grains diffusing outwards from the cold region at smaller radii both deposit their vapour in the hot region, causing the vapour density there to rapidly increase. As the vapour pressure increases, more vapour can condense onto the grains at the same temperature; as this condensation commences, the opacity once again starts to increase, causing the disc to cool. The cooling and condensation then enter a runaway phase, and the disc returns to the cold state.}

In the region that enters the limit-cycle, we can construct S-curves in temperature and mass-flux space; these are shown in the left and right panels of Fig.~\ref{scurve_isoT}. S-curves are used to plot out the contours of stable solutions in unstable systems; we see how entering a limit-cycle causes the system to traverse the stable solution space, ultimately restarting once one period of the cycle has completed. For these S-curves, we see that the disc is initially in the cold, high mass-flux state; after $\sim20$~kyr the disc heats up enough to jump to the hot, low mass-flux state, where it remains for a few thousand years before jumping back to the cold state. In contrast to the S-curves shown by \cite{owen2020}, the CO surface density continues to increase once the disc has jumped back to the cold state. This is because the ice is initially deposited onto the smallest grains, and it takes several thousand years at this radius for the ice to be transferred to larger grains that can increase the CO mass-flux to greater values than the mass-flux that is feeding the region from the outer disc. This can be seen in how the ice-density-averaged Stokes number changes over the cycle. Once the mass-flux surpasses the feeding rate, we see that the surface density and mass-flux begin to decrease, continuing the cycle.

\subsection{Full 2D solutions}

\begin{figure}
    \centering
    \includegraphics[width=0.99\columnwidth, trim={0.2cm 0.1cm 0.2cm 0.1cm}, clip]{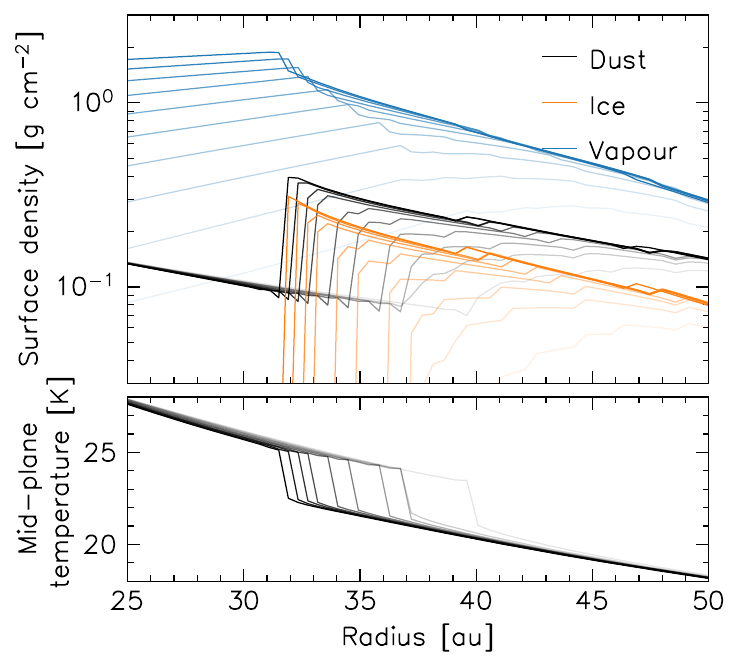}
    \caption[CO surface density]{Evolution of the fiducial full 2D simulation. The simulation has run for $\sim4$~Myr; line opacity increases with time in intervals of 0.4~Myr. The top row shows the surface densities of dust (ice included), ice and vapour, whilst the second row shows the mid-plane temperature. The small non-continuous deviations in the surface density appear due to integration over the tightly-constrained, discontinuous vertical structure of the ice.}
    \label{sigmas_nonisoT}
\end{figure}

\begin{figure}
    \centering
    \includegraphics[width=0.99\columnwidth, trim={0.5cm 0.5cm 0.5cm 0.5cm}, clip]{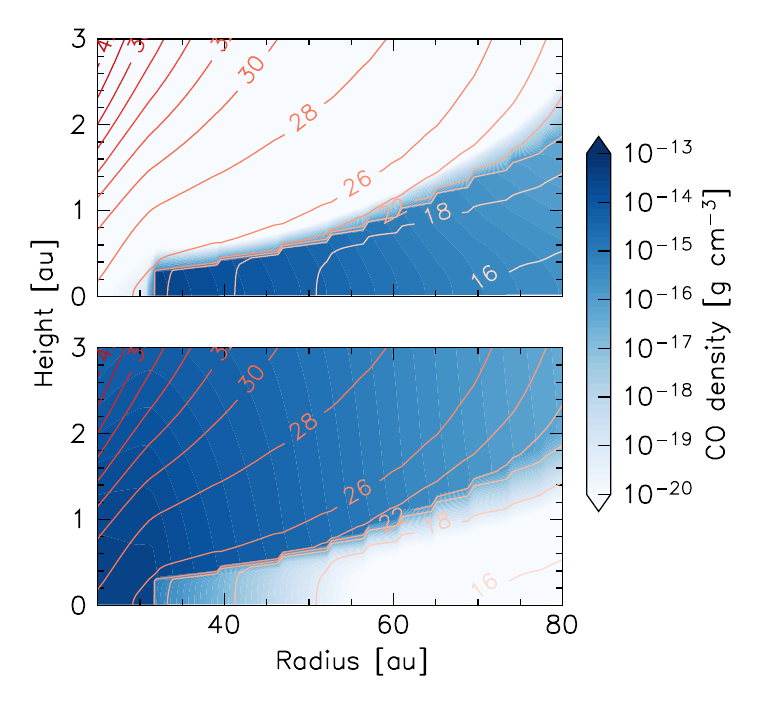}
    \caption[Ice and vapour density]{Spatial distributions of CO ice and vapour density in the full 2D fiducial simulation. Temperature contours (in K) are over-plotted for reference.}
    \label{icevapdens}
\end{figure}

We now run full 2D calculations, where the ice-vapour solver is given the non-vertically-isothermal temperature structure. As our fiducial simulation, we repeat the vertically-isothermal simulation presented in Section~\ref{vertisorun}, which has the parameters: $\alpha=10^{-3}$, $v_\text{frag}=10~$\unit{\metre\per\second}, $M_\text{disc}=0.05~$M$_\odot$, and outer disc values of the ice mass-fraction and dust-to-gas ratio of 20\% and 1\% respectively. Allowing the system to evolve for several Myr, we find that the disc does not enter a limit-cycle, instead evolving very slowly towards equilibrium in the cold state. This can be seen in the surface density profiles shown in Fig.~\ref{sigmas_nonisoT}, where the snow line can be seen to move radially-inwards without ever entering a limit-cycle. The vapour dominates the total volatile surface density, even 10s of au exterior to the mid-plane snow line. As can be seen in Fig.~\ref{icevapdens}, including the vertical temperature structure tightly constrains the snow surface to low altitudes above the mid-plane. Above the snow surface, there is a large enough reservoir of vapour that the ice is only dominant over vapour outside of $\sim80~$au --- this is in contrast to the vertically-isothermal model (and the model presented in \citealt{owen2020}) where the ice dominates the vapour at all radii exterior to the mid-plane snow line radius. 

\begin{figure}
    \centering
    \includegraphics[width=0.99\columnwidth, trim={0.25cm 0.1cm 0.1cm 0.1cm}, clip]{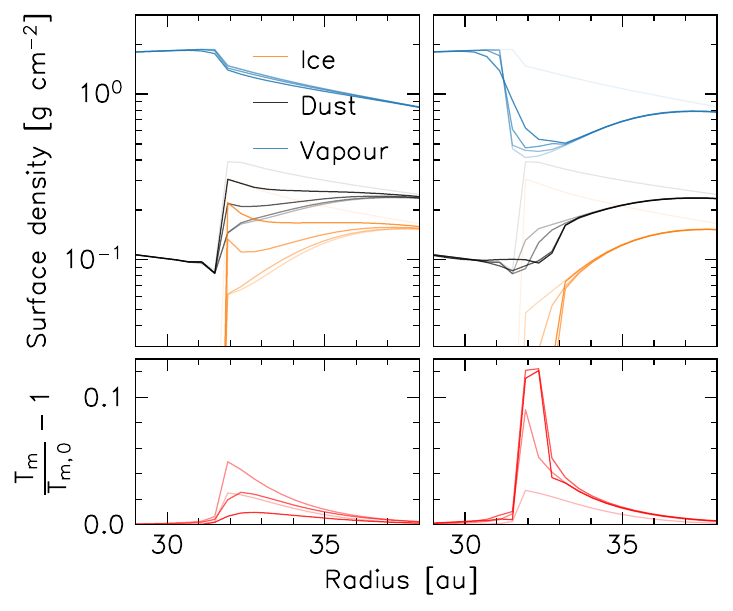}
    \caption{Evolution of the fiducial full 2D simulation after perturbation. The left column shows the evolution after a perturbation to the ice density, whilst the right column shows the evolution after a perturbation to the total volatile density. Line opacity increases with time, with snapshots at $t=0$, 4, 27, 176 and 1150~yr. The top row shows the surface densities of dust (ice included), ice and vapour, whilst the second row shows the change in mid-plane temperature compared to the unperturbed state.}
    \label{perts}
\end{figure}

To investigate what is stopping the disc from entering a limit-cycle, we perturb the volatile density in the vicinity of the mid-plane snow line and study the subsequent changes in temperature and volatile density. To perturb the volatile density, we apply a Gaussian decrease centred on the mid-plane snow line radius that has an amplitude of 80\% of the initial density and a standard deviation of 3~au. We present two cases in Fig.~\ref{perts}; the left column shows the subsequent evolution after a perturbation to solely the ice density, whilst the right column shows the subsequent evolution after a perturbation to both the ice and vapour components. In both cases, we see that the mid-plane temperature initially begins to rise. This is in line with our expectations for the system if it is unstable, as a decrease in the ice density should lead to an increase in the temperature through making the cooling less efficient. However, after this initial temperature increase, the cases diverge. In the case with only a perturbation to the ice component, the ice surface density does not continue to decrease in tandem with the thermal perturbation; instead, it gradually begins to increase back to the initial distribution. This evolution back to equilibrium is also then seen in the temperature structure, which cools back down. On the other hand, the case where the vapour is also perturbed sees the instability develop as expected; the thermal perturbation at the mid-plane snowline leads to a further decrease in ice density, which continues to raise the temperature until all of the ice has sublimated at the initial snow line location. This pushes the mid-plane snow line back to a radius at which the initial perturbation was not large enough to trigger sufficient mid-plane heating.

\begin{figure}
    \centering
    \includegraphics[width=0.99\columnwidth, trim={0.25cm 0.2cm 0.2cm 0.2cm}, clip]{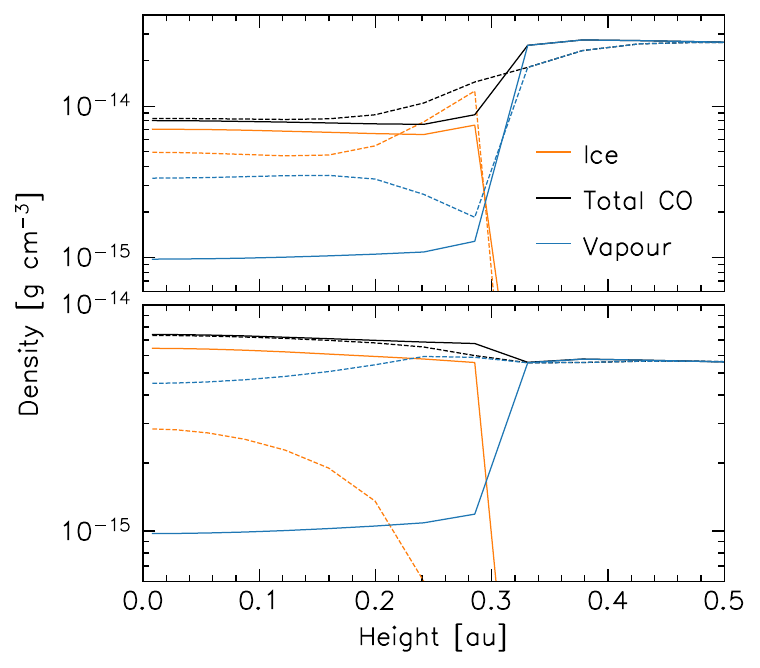}
    \caption{Vertical profiles of the ice and vapour densities for each of the perturbation cases. The ice-only and total volatile cases are shown in the top and bottom panels respectively. The solid lines show the systems just after perturbation, whilst the dashed lines show the systems after 17~yr of evolution.}
    \label{pert_zdens}
\end{figure}

The reason that the unperturbed vapour reservoir is able to damp the development of the instability is due to vertical transport of vapour over the 2D snow surface. Fig.~\ref{pert_zdens} shows the evolution of the vertical density distributions of the volatile components for both perturbation cases. In the ice-only case, we see that the initial perturbation generates a large gradient in the total volatile density distribution. This drives vapour to diffuse vertically-downwards, condensing onto the dust grains beneath the snow surface. This supply of ice is able to counteract the sublimation driven by the thermal perturbation. This behaviour is in contrast to the case where the vapour is also perturbed, where the lack of an large volatile gradient after the initial perturbation means that the diffusive flux of vapour is significantly diminished, and the ice sublimates faster than it is replenished --- allowing the instability to grow.

\subsubsection{Parameter exploration}

\begin{figure}
    \centering
    \includegraphics[width=0.99\columnwidth, trim={0.5cm 0.5cm 0.5cm 0.5cm}, clip]{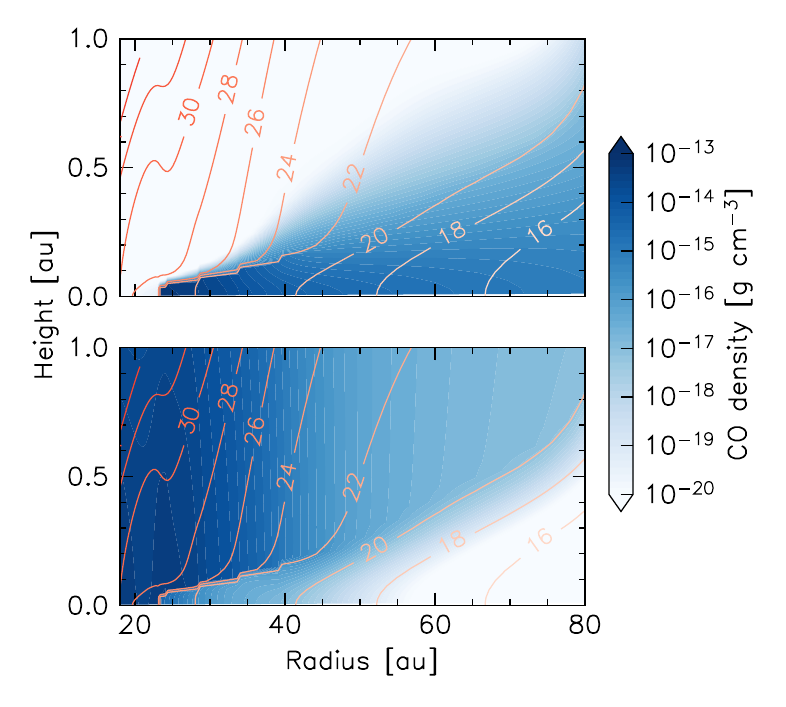}
    \caption[Ice and vapour density]{Spatial distributions of CO ice and vapour density in the full 2D simulation with lower turbulence ($\alpha=10^{-4}$). Temperature contours (in K) are over-plotted for reference.}
    \label{icevapdens_a1e-4}
\end{figure}

To explore whether this behaviour is limited to our choice of fiducial parameters, we run a number of simulations with the key parameters varied. First we vary the turbulence level, running a simulation with a reduced turbulence of $\alpha=10^{-4}$. Reducing the turbulence level decreases the diffusive flux of vapour, making the diffusion time longer. This affects the vertical transport of vapour, and could therefore dictate whether the disc can be unstable. However, we once again see a stable disc configuration with a dominant vapour reservoir throughout much of the disc extent. Fig.~\ref{icevapdens_a1e-4} shows the 2D spatial distributions of ice and vapour after 2~Myr for this low turbulence simulation. Compared to the fiducial simulation, the snow surface lies closer to the mid-plane. This is because the temperature rises more rapidly away from the mid-plane, because the lower turbulence level leads to enhanced dust settling that brings the photospheric surface of the disc to lower altitudes. Even though the lower turbulence increases the diffusion timescale of the vapour, the tightly-constrained snow surface allows such a large vapour reservoir to exist that it is still able to supply the mid-plane with enough ice through vertical transport to quench the instability.

\begin{figure}
    \centering
    \includegraphics[width=0.99\columnwidth, trim={0.2cm 0.2cm 0.2cm 0.2cm}, clip]{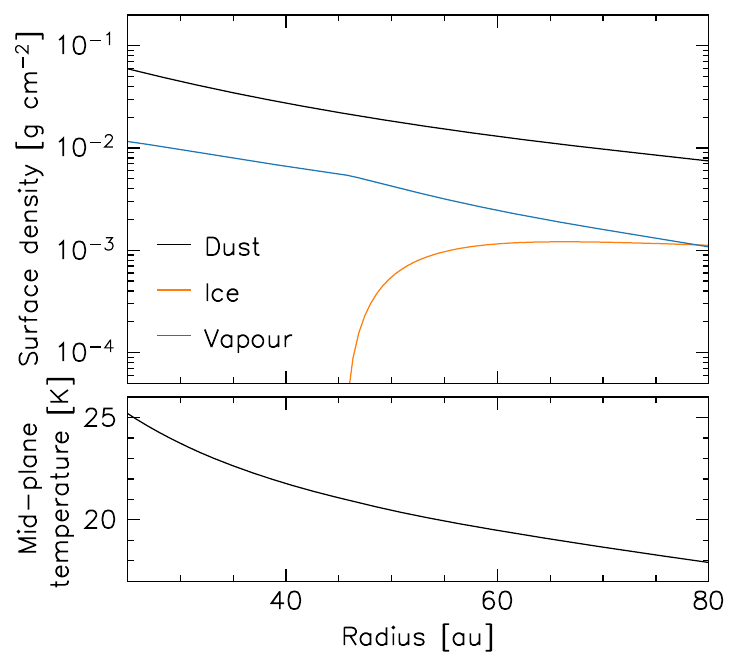}
    \caption{The steady-state reached for the full 2D simulation with a lower fragmentation velocity ($v_\text{frag}=1$~\unit{\metre\per\second}). The top row shows the surface densities of dust (ice included), ice and vapour, whilst the second row shows the mid-plane temperature. }
    \label{v100_sigmas}
\end{figure}

We also vary the fragmentation velocity of the grains,  running a simulation with $v_\text{frag}=1$~\unit{\metre\per\second}. A lower fragmentation velocity leads to a grain-size distribution with a larger proportion of small grains, so we expect the snow surface to be less tightly-constrained to the mid-plane due to a higher altitude photosphere. This could affect the instability by reducing the size of the vapour reservoir above the snow surface that is available to supply the mid-plane. However, we once again find a stable disc configuration. Fig.~\ref{v100_sigmas} shows the surface density distributions of ice, dust and vapour after 1~Myr for a simulation with the fiducial turbulence but a lower fragmentation velocity. Whilst the lower fragmentation velocity leads to a less-constrained snow surface as expected, the low Stokes numbers of the ice-carrying grains severely restrain the ice mass-flux. This reduces the dynamical separation between the ice and vapour phase volatiles, meaning that vapour does not build up at the snow surface. With no build-up of vapour, there is no accruing of ice behind the snow line due to vapour diffusion meaning that the mass-flux conditions required to trigger the instability cannot be established. We therefore find that the disc reaches a steady-state.

\section{Discussion}
\begin{figure*}
    \centering
    \includegraphics[width=0.99\textwidth, trim={0.25cm 0.1cm 0.1cm 0.1cm}, clip]{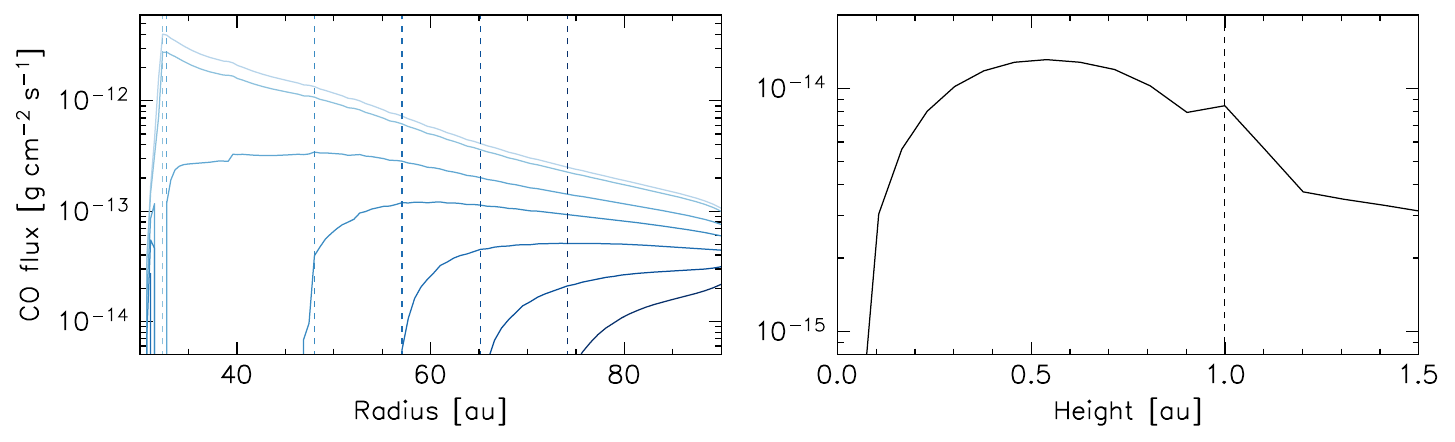}
    \caption{The left panel shows the radial CO flux at a range of heights (height increases with line darkness); 0, 0.13, 0.3, 0.57, 0.85, 1.2 and 1.6~au. The flux is defined as positive in the radially-inward direction, i.e. towards the star. The right panel shows the the vertical CO flux at a radius of 60~au. The flux is defined as positive in the vertically-downward direction, i.e. towards the mid-plane. The vertical dashed lines correspond to the snow line locations at each height (left panel) or radius (right panel).}
    \label{COflux_nonisoT}
\end{figure*}

The perturbation tests show that the disc can be unstable and would be able to enter a limit-cycle if the vapour reservoir was unable to supply the mid-plane. However, a large vapour reservoir is a natural consequence of the 2D structure of the snow surface. The tightly-constrained shape means that the snow line radius shifts radially-outwards as a function of height above the mid-plane, and vapour is therefore deposited at all radii, not solely at the mid-plane snow line radius. Due to this, the disc cannot enter a limit-cycle, as the CO mass-flux conditions required to facilitate the cycle seen in the vertically-isothermal simulations are prohibited. We can see this by examining the radial and vertical fluxes of CO of the unperturbed system. The left panel of Fig.~\ref{COflux_nonisoT} shows how the radial flux of CO, i.e. 

\begin{equation}
    \label{COflux}
    F_R = \rho_\text{i} (v_{\text{i},R}^\text{diff} + v_{\text{i},R}^\text{adv})+ \rho_\text{v} (v_{\text{v},R}^\text{diff} + v_{\text{v},R}^\text{adv}),
\end{equation} 
where both the diffusive and advective velocities of the ice and vapour are included, varies as a function of height for the final snapshot shown in Fig.~\ref{sigmas_nonisoT}. Around the mid-plane, we see that the flux has a similar radial profile to that seen in the vertically-isothermal model; for radii exterior to the mid-plane snow line there is a negative slope, which we may expect to lead to a decrease in CO density as a function of time. This decrease is countered by the deposition of vapour in this region at higher altitudes in the disc, as shown by the mass-fluxes at heights where the snow line radii fall in this region. The vapour deposited at these heights is transported in the vertical direction through diffusion, crossing the snow surface and feeding the mid-plane with CO ice. This is demonstrated by the vertical flux (defined similarly to the flux in Eqn.~\ref{COflux}, but with the $Z$-velocities) at 60 au shown in the right panel of Fig.~\ref{COflux_nonisoT}. The vertical flux is defined as positive towards the mid-plane, and we can therefore see that there is a net flux of volatiles above the snow surface heading towards the mid-plane. The flux decreases as we approach the mid-plane, meaning that the density of volatiles would increase over time were there not also radial fluxes to consider. This shows that the vertical transport of volatiles is able to supply the mid-plane and counter radial fluxes that would lead to the disc entering a limit-cycle. In the vertically-isothermal model, there is no vapour reservoir able to vertically resupply the mid-plane, allowing the formation of a limit-cycle. We make a note here that we do not consider vertical gas flows due to photoevaporative or MHD winds, which one could expect to affect how well the vapour can resupply the mid-plane. However, our snow surfaces are highly constrained to the mid-plane, where large gas densities mean that upward velocities due to winds launched near the disc surface are small. This means any advective flux of vapour away from the mid-plane would be much smaller than the diffusive fluxes driven by the large vapour gradient at the snow surface. We are therefore confident that our results would be unaffected by the inclusion of wind-driven vertical velocities.

\cite{owen2020} speculated on the effect that shadowing due to features in the dust density distribution would have on the instability, suggesting that it could spawn more time-variable features. However, in both the vertically-isothermal and full 2D models, whilst the photosphere shapes do indicate a shadowed region exterior to the mid-plane snow line radius, the extent of this shadowing is not as extreme as suggested by \cite{owen2020} and therefore no additional features appear in the outer disc.

Our findings indicate that snow lines in protoplanetary discs can be thermally unstable, with two stable equilibrium solutions. However, unlike the results of \cite{owen2020}, the disc cannot enter a limit-cycle that causes it to jump between the solution branches in an oscillatory fashion with a regular period. We therefore infer that the snow line instability is unlikely to be the cause of any structures observed in disc surveys. We expect that the only way in which the instability may manifest in realistic discs is via stochastic events where the opacity and mass-flux conditions are such that the disc is forced to jump between states. One speculative example could be a gap opening due to a planet or other process that traps dust and ice exterior it, generating a build-up of volatiles that can trigger the disc to jump to the cold state. This is something to bear in mind for future studies; if significant transient dynamical evolution is seen in simulations of snow lines in 2D, thermodynamically-evolving discs, one can test whether it is caused by stochastic triggering of the instability by checking for optical depths in the range $0.1-0.5$ and how the temperature evolves as the volatile density distribution evolves. We also make the point that a limit-cycle could appear if some disc configuration that we have not studied leads to a large vertically-isothermal region in the vicinity of the snow line.

In addition, we have shown that including the vertical structure is vitally important for correctly modelling the evolution of volatiles in protoplanetary discs. As the snow line radius is strongly dependent on altitude, the spatial distribution of volatiles cannot be correctly evolved by assuming that vapour builds up solely at the radius of the mid-plane snow line. Instead, the vapour is deposited over a wide range of radii, and the vertical transport of this vapour means that the ice distribution is also affected across this radial range. This could have an impact on the results of 1D models of volatile evolution in discs, where build-ups of volatiles are found at the mid-plane radii of snow lines due to the ``cold-finger'' effect \cite[see e.g.][]{stevenson1988,cuzzi2004,schoonenberg2017}. Also, the vapour reservoir that exists above the snow surface needs to be considered if running 1D models due to its impact on the spatial (especially vertical) variation of quantities important to planet formation, such as the C/O ratio.

Our results have also shown that the mid-plane snow line radius and the 2D structure of the snow surface are highly dependent on the spatial structure and grain-size distribution of the dust due to how these govern the thermal structure of the disc. Both of these features of the dust component can be modulated by dust microphysical parameters, such as the fragmentation velocity, or the nature of the gas evolution, such as the level of turbulence. As we know that the dust component evolves dramatically over the several million years that discs exist, we would similarly expect the snow line structure to evolve. This has implications for planet formation, where the locations of the snow lines of different species govern the composition of icy pebbles or gas that a nascent planet accretes as it grows and migrates through the disc. Future work will be to investigate how snow lines evolve in discs with evolving dust populations by running simulations with no fixed mass-flux feeding the outer boundary as was done for this work.

\section{Conclusions}

In this paper, we have revisited the snow line instability presented in \cite{owen2020}, using the disc evolution code \textsc{cuDisc} with a newly-added ice-vapour chemistry module. The added ice-vapour chemistry includes time-dependent calculation of adsorption and desorption across a full grain-size distribution, collisional evolution of the icy grains, and coupling of the ice to the disc thermodynamics through modulation of the dust opacities as the ice mass-fraction changes. When investigating the snow line instability, we find that:

\begin{itemize}
    \item From 2D equilibrium models of the CO snow line and snow surface, we find that the disc is in fact unstable when including the vertical dimension, as two stable solutions exist. These two solutions manifest as two possible locations for the snow line in the disc; the solution where the snow line lies closer to the star is the cold state, whilst the solution where the snow line lies further out is the hot state.
    \item However, in dynamic models, we find that the disc can only be made to periodically jump between the two solutions (i.e. enter a limit-cycle as seen by \citealt{owen2020}) in models where the ice-vapour solver is fed a vertically-isothermal temperature structure, where all altitudes have the mid-plane temperature.
    \item Simulations with the full, non-isothermal vertical temperature structure do not exhibit a limit-cycle, due to being stabilised by the presence of a vapour reservoir above the snow surface that can feed volatiles to the mid-plane through vertical diffusive transport.
    \item This vapour reservoir is a natural outcome of the 2D temperature structure, which increases away from the mid-plane and leads to a snow surface that is tightly constrained to the disc mid-plane. As icy grains drift inwards, they therefore cross the snow surface over a wide range of radii, depositing vapour throughout a large radial extent of the disc.
    \item These results have shown that including the vertical structure is vital for correctly modelling the dynamical evolution of volatiles in discs, particularly around mid-plane snow lines.
    \item Snow surfaces are exponentially sensitive to the disc thermal structure and therefore the nature of the dust and gas; we expect their mid-plane location and vertical structure to evolve substantially over the disc lifetime.
\end{itemize}

In future work, we will study how snow lines evolve in thermodynamically-evolving discs over Myr timescales for a range of disc parameters with no fixed-outer mass-flux; this will allow us to get a handle on how the shape and structure of 2D snow surfaces can vary over the disc lifetime, which may have a profound effect on our understanding of planet formation.

\section*{Acknowledgements}

This project has received funding from the European Research Council (ERC) under the European Union’s Horizon 2020 research and innovation programme (Grant agreement No. 853022). AR is supported by Science and Technology Facilities Council (STFC) award ``Building Rocky Planets in the Inner Disc: From Dust to Planetesimals" (project reference UKRI1189). JEO and RAB are supported by Royal Society University Research Fellowships. Some of this work used the Distributed Research using Advanced Computing (DiRAC) Data Intensive service (Cambridge Service for Data Driven Discovery, CSD3) at the University of Cambridge, managed by the University of Cambridge University Information Services on behalf of the STFC DiRAC High Performance Computing (HPC) Facility (\url{www.dirac.ac.uk}). The DiRAC component of CSD3 at Cambridge was funded by Department for Business, Energy \& Industrial Strategy (BEIS), UK Research and Innovation (UKRI) and STFC capital funding and STFC operations grants. DiRAC is part of the UKRI Digital Research Infrastructure.

\section*{Data Availability}

The GitHub repository for \textsc{cuDisc} can be found at \url{https://github.com/cuDisc/cuDisc/}, which includes the new developments described here. Please feel free to contact Alfie Robinson (a.robinson21@imperial.ac.uk) for any queries relating to the work described in this paper. Data will be made available on reasonable request.



\bibliographystyle{mnras}
\bibliography{refs} 




\appendix

\section{Comparison of opacity model to Mie theory}

\begin{figure}
    \centering
    \includegraphics[width=\columnwidth,trim={0.2cm .2cm 0.2cm 0.2cm},clip]{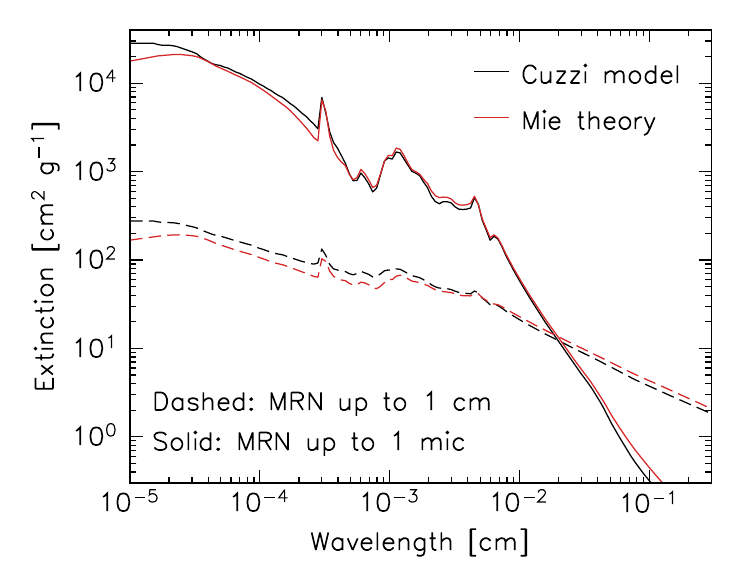}
    \caption{Comparison of our opacity model to Mie theory for the DSHARP composition averaged over two grain-size distributions.}
    \label{cuzzimiecomp}
\end{figure}

Fig.~\ref{cuzzimiecomp} shows how extinctions ($\kappa_a+\kappa_s$) calculated using our opacity model (based on that of \citealt{cuzzi2014}) compare to full Mie theory. For the comparison we have used the DSHARP composition and calculated the Mie opacities using the DSHARP Python opacity tool \citep{DSHARP2018}. Extinctions averaged over two MRN grain-size distributions are shown: one of small dust with a maximum grain size of 1~\unit{\micron}, and one of a wide range of sizes with a maximum of 1~cm. The extinctions are mostly very consistent, with the largest deviations ($\sim50$\%) occurring only at the very shortest wavelengths; these wavelengths lie at shorter wavelengths than those corresponding to the peak of a typical proto-stellar spectrum ($\sim0.7$~\unit{\micron}) so will not lead to large deviations when considering the absorption/scattering of the majority of stellar irradiation.

\section{1D method}\label{1Dmethods}

For our 1D model of dust evolution, we solve the 1D advection-diffusion equation

\begin{equation}
    \label{dust1D}
    \frac{\partial \Sigma_\text{d}}{\partial t} + \frac{1}{R} \frac{\partial}{\partial R} \left\{R \left[ \Sigma_\text{d} v_{\text{d},R} - \frac{\nu \Sigma_\text{g}}{\text{Sc}} \frac{\partial}{\partial R} \left(\frac{\Sigma_\text{d}}{\Sigma_\text{g}} \right) \right] \right\} = 0,
\end{equation}
using the same second-order finite-volume Godonuv method used to solve the 2D advection-diffusion equations \cite[see section 3 in][]{robinson2024}.
The radial advection velocity for a particular grain-size is calculated via 

\begin{equation}
    \label{1DvRdust}
    v_{\text{d},R} = \frac{\bar{v}_{\text{g},R} + 2(v_{\text{g},\phi}-v_\text{K}\text{St}) }{1+\text{St}^2},
\end{equation}
where $\bar{v}_{\text{g},R}$ is the radial gas velocity vertically-integrated over the dust distribution for that given grain-size,
\begin{multline}
    \label{vgbar}
    \bar{v}_{\text{g},R} = \frac{1}{\Sigma_\text{d}}\int^\infty_{-\infty}\rho_\text{d} v_{\text{g},R} \text{d}Z \\
    =  \frac{\sqrt{1+\text{St}/\alpha}}{\Sigma_\text{g}} \int_{-\infty}^{\infty}\rho_\text{g} v_{\text{g},R} \exp\left(-\frac{Z^2}{2H_\text{d}^2}\frac{\text{St}}{\alpha}\right) \text{d}Z,
\end{multline}
where we have made use of the diffusion-settling dust scale-height given by $H_\text{d} = H \sqrt{\frac{1}{1+\text{St}/\alpha}}$, and $v_{\text{g},\phi}$ is the 1D gas azimuthal velocity.

\bsp	
\label{lastpage}
\end{document}